\documentclass[a4paper,fleqn,usenatbib]{mnras}

\usepackage{mathptmx}
\usepackage{txfonts}

\usepackage[T1]{fontenc}
\usepackage{ae,aecompl}

\usepackage{graphicx}	

\usepackage{longtable}

\title[SU Lyn - a transient SySt]{SU Lyn - a transient symbiotic star}

\author[I{\l}kiewicz et al.]{Krystian I{\l}kiewicz,$^{1}$\thanks{E-mail: krystian.a.ilkiewicz@durham.ac.uk}
Joanna Miko{\l}ajewska,$^{2}$
Simone Scaringi,$^{1}$
Fran\c cois Teyssier,$^{3}$
\newauthor
Kiril A. Stoyanov,$^{4}$ and
Matteo Fratta$^{1}$
\\
$^{1}$Centre for Extragalactic Astronomy, Department of Physics, University of Durham, South Road, Durham, DH1 3LE, UK\\
$^{2}$ Nicolaus Copernicus Astronomical Center, Polish Academy of Sciences, Bartycka 18, 00716 Warsaw, Poland\\
$^{3}$Observatoire Rouen Sud, 67 rue Jacques Daviel, 76100 Rouen, France\\
$^{4}$Institute of Astronomy and National Astronomical Observatory, Bulgarian Academy of Sciences, Tsarigradsko Shose 72, BG-1784 Sofia, Bulgaria
}

\date{Accepted XXX. Received YYY; in original form ZZZ}

\pubyear{2020}

\begin{document}
\label{firstpage}
\pagerange{\pageref{firstpage}--\pageref{lastpage}}
\maketitle

\begin{abstract}
SU Lyn is a binary system composed of a white dwarf and a red giant star.  Although it is known to be bright and variable at X-ray wavelengths, the optical counterpart of the source appeared as a single red giant without prominent emission lines. Because of the lack of optical features typical for interacting systems, the system was classified as a hidden symbiotic star. We present the results of optical monitoring of the system. While SU~Lyn did not show substantial photometric variability, the spectroscopic observations revealed a complex behavior. The system showed strong emission line variability, including P~Cygni profiles, changing line emission environments, and variable reddening. Both X-ray and optical observations indicate that the components of SU Lyn were interacting only for a short time during the last twelve years of monitoring. For the first time we showed that SU~Lyn resembled a classical symbiotic star when it was X-ray bright, and remained hidden afterwards. We also discuss the current evolutionary status of the red giant, as well as possible future evolution of the system. We suggest that SU~Lyn could be a progenitor of a classical, persistent symbiotic system.
\end{abstract}




\section{Introduction}

Symbiotic stars are interacting binaries with red giant donors. The mass accretor is typically a white dwarf, while in some systems a neutron star is present instead of the white dwarf. Their orbital periods range from hundreds of days to decades \citep{2009AcA....59..169G,2013AcA....63..405G}. The circumstellar material in symbiotic stars is ionized by the mass accretor, which leads to the production of emission lines. The main source of energy in symbiotic stars is often the accretion itself. However, in some systems the white dwarf experiences steady shell-burning, which is providing more radiation energy ionizing the circumbinary material, resulting in more prominent emission lines..

SU~Lyn was discovered as a semi-regular variable star with a period of $\sim$126\,days by \citet{1955AN....282...73K}. The spectral type of the giant in SU~Lyn was first estimated to be M2 \citep{1956PMcCO..13a...0B}. Later, the spectral type was derived to be M6 \citep{1983ApJS...53..413G} and finally it was refined to be between M5.6 and M5.9 \citep{2016MNRAS.461L...1M}. Based on optical and infrared magnitudes the red giant in SU~Lyn was estimated to have a diameter of $\sim$3.5~mas \citep{2014ASPC..485..223B}. No water maser emission was detected from SU~Lyn \citep{1995A&A...296..727S}. 

 \citet{2016MNRAS.461L...1M} identified SU~Lyn with a hard X-ray source  4PBC J0642.9+5528 detected by the \textit{Swift} satellite. In addition to X-ray radiation, \citet{2016MNRAS.461L...1M} detected UV variability, weak Balmer, \mbox{[Ne\,{\sc iii}]} and \mbox{Ca\,{\sc ii}} emission lines, and red giant optical spectrum. Based on X-ray and UV variability, \citet{2016MNRAS.461L...1M} classifies SU~Lyn as a symbiotic star that is not showing strong emission lines typical for this class of objects. They also argued that SU~Lyn does not show strong emission lines because it is not a shell-burning system. The symbiotic star nature of SU~Lyn was later supported by the discovery of high-excitation emission lines in SU~Lyn UV spectrum \citep{2021MNRAS.500L..12K}. A detailed review of SU~Lyn optical properties is given by \citet{2021arXiv210402686M}.  Because SU~Lyn would not be detected as a symbiotic star in optical surveys \citet{2016MNRAS.461L...1M}  suggested that this system represents a large hidden population of symbiotic stars. These stars would not show typical symbiotic features in the optical spectrum similar to SU~Lyn, and there is no estimate of the size of their population. However, the infrared colors of SU~Lyn seem to match the symbiotic star population \citep{2019MNRAS.483.5077A}.

SU~Lyn showed strong variability in X-ray and UV bands \citep{2016MNRAS.461L...1M,2018ApJ...864...46L,2021MNRAS.500L..12K}. \citet{2018ApJ...864...46L} showed that SU~Lyn harbors a relatively massive (>0.7\,M$_\odot$) white dwarf. The mass transfer rate in the system decreased by nearly 90\% between 2015 and 2016 \citep{2018ApJ...864...46L}.  Moreover, \citet{2018ApJ...864...46L} showed that there is a strong, variable source of absorption in the system and that the boundary layer of the white dwarf became more optically thin with time.

Here we show results of long optical monitoring of SU~Lyn, study the evolutionary status of the red giant as well as discuss the nature of the system. In particular, we show that SU~Lyn optical spectrum resembled that of a classical symbiotic system when the system was X-ray bright. We argue that the SU~Lyn can evolve into a classical, stable symbiotic star.

\section{Observations}

\subsection{Photometry}

\begin{figure*}
	\includegraphics[width=\textwidth]{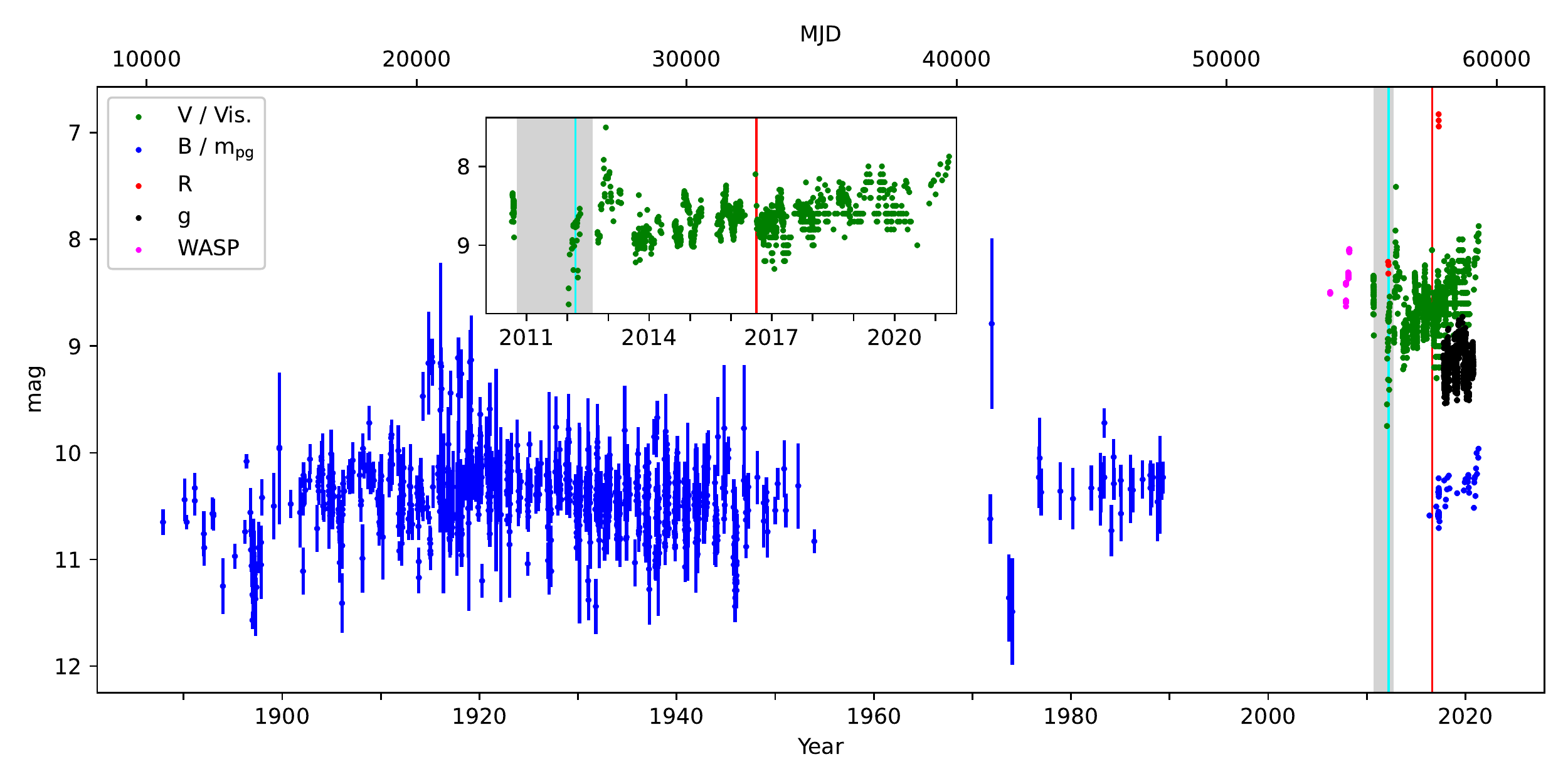}
    \caption{Long-term light curve of SU~Lyn. The cyan line marks the date of SOPHIE spectrum. The red line marks the observation of \citet{2018ApJ...864...46L}. The gray area marks the X-ray high state observed by \citet{2016MNRAS.461L...1M}.}
    \label{fig:lc_hist}
\end{figure*}

We collected all of the photometry available in the literature for SU~Lyn. Archival data from the Harvard College Observatory in the photographic m$_{\mathrm{pg}}$ filter were collected from DASH \citep[Digital Access to a Sky Century at Harvard;][]{2010AJ....140.1062L}. Optical observations in a proprietary filter were collected from the Wide Angle Search for Planets ($WASP$) project \citep{2010A&A...520L..10B}. Data in the standard optical filters were collected from the International Gamma Ray Astrophysics Laboratory ($INTEGRAL$) Optical Monitoring Camera \citep[OMC;][]{2003A&A...411L.261M}, All-Sky Automated Survey for Supernovae \citep[ASAS-SN;][]{2014ApJ...788...48S,2017PASP..129j4502K} and American Association of Variable Star Observers (AAVSO) database. The available photometric data ranged from years 1887 to 2021, with major gaps in 1960's and from years 1989 to 2006. The collected light curve of SU~Lyn is presented in Fig.~\ref{fig:lc_hist}.

\subsection{Spectroscopy}

\begin{figure}
\resizebox{\hsize}{!}{\includegraphics{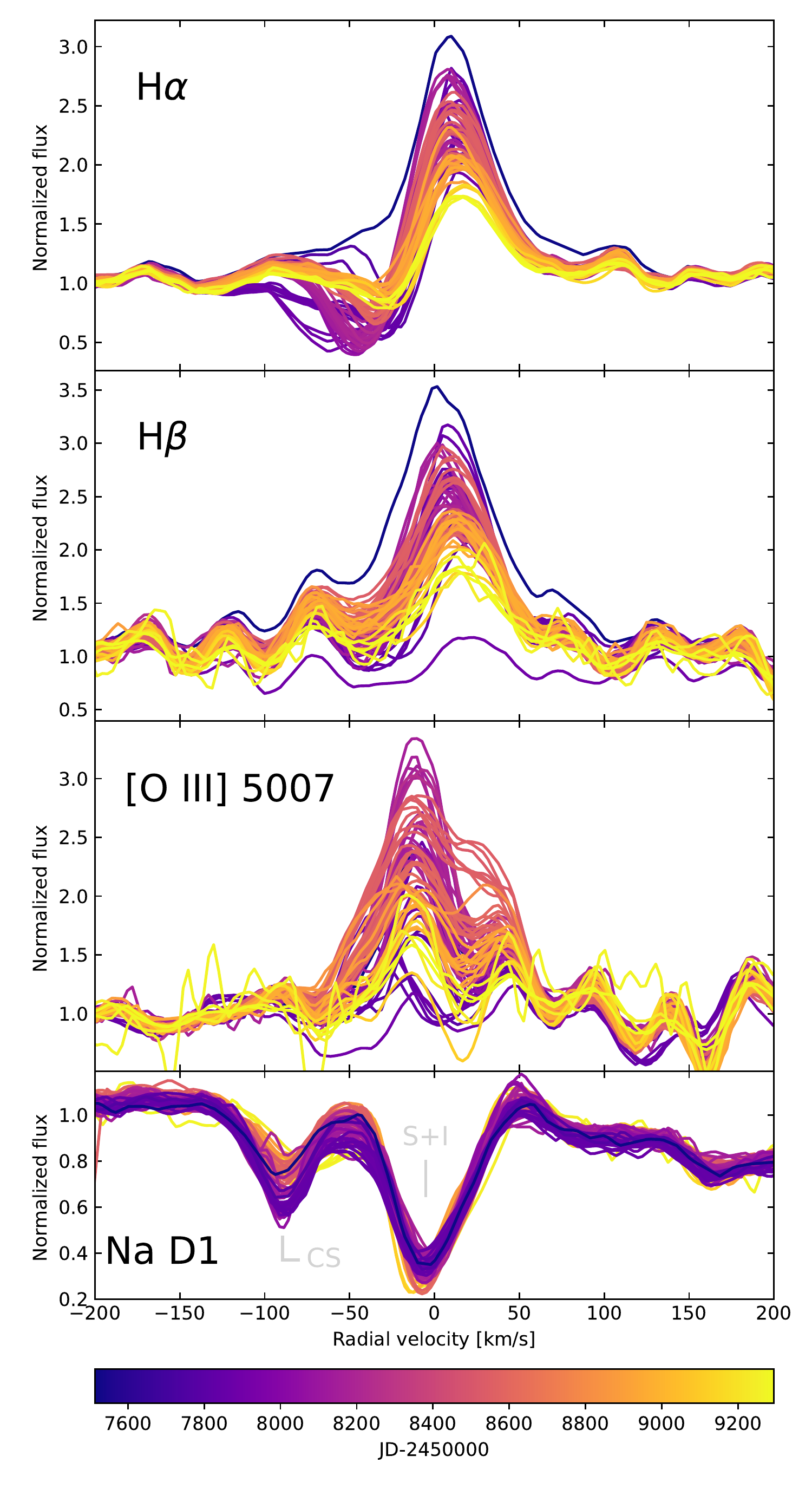}}
    \caption{Emission lines in ARAS spectra after normalization, subtracting a normalized spectrum of 13~Lyr as well as Na~D1 absorption line for witch no subtracting was performed (see text). All the spectra were  shifted by the system mean radial velocity. The emission line region spectra were shifted by +1 in the flux scale for clarity. SOPHIE spectrum was not plotted. The non-variable combination of stellar and interstellar profiles of Na~D1 line is denoted by ''S+I'' and a variable circumstellar component of Na~D1 line is denoted as ''CS''.}
    \label{fig:all_spec}
\end{figure}

We collected spectroscopic observations from the  Astronomical Ring for Access to Spectroscopy database\footnote{https://aras-database.github.io/database/symbiotics.html} (ARAS; \citealt{2019CoSka..49..217T}). The low-resolution spectra (R$<$10000) of SU~Lyn do not show any significant changes compared to the spectra presented in the literature. Hence, we analyzed only high-resolution spectra available in ARAS. The list of observations is presented in Table~\ref{logspec}. All the spectra were reduced by the individual observers following standard procedures. Since the spectra from ARAS were carried out using different instruments, we normalized all of the spectra to the local continuum in order to avoid inconsistent spectrophotometric calibration. As the emission lines in SU~Lyn are relatively weak they can be hidden within the red giant absorption spectrum. To improve their visibility we subtracted from ARAS spectra a spectrum of a M5\,III star spectroscopic standard. For the spectroscopic standard spectrum we chose a spectrum of 13~Lyr carried out by Fran\c cois Teyssier that was shifted to the SU~Lyn radial velocity. For the radial velocity, we used the mean radial velocity of the red giant (Section~\ref{rvsearch}).  We did not subtract the 13~Lyr spectrum from the region near the Na~D1 line spectra region to avoid contamination coming from Na~D1 lines originating from the interstellar reddening of 13~Lyr. The emission lines and Na~D1 absorption lines are presented collectively in Fig.~\ref{fig:all_spec}.

\begin{figure*}
	\includegraphics[width=\textwidth]{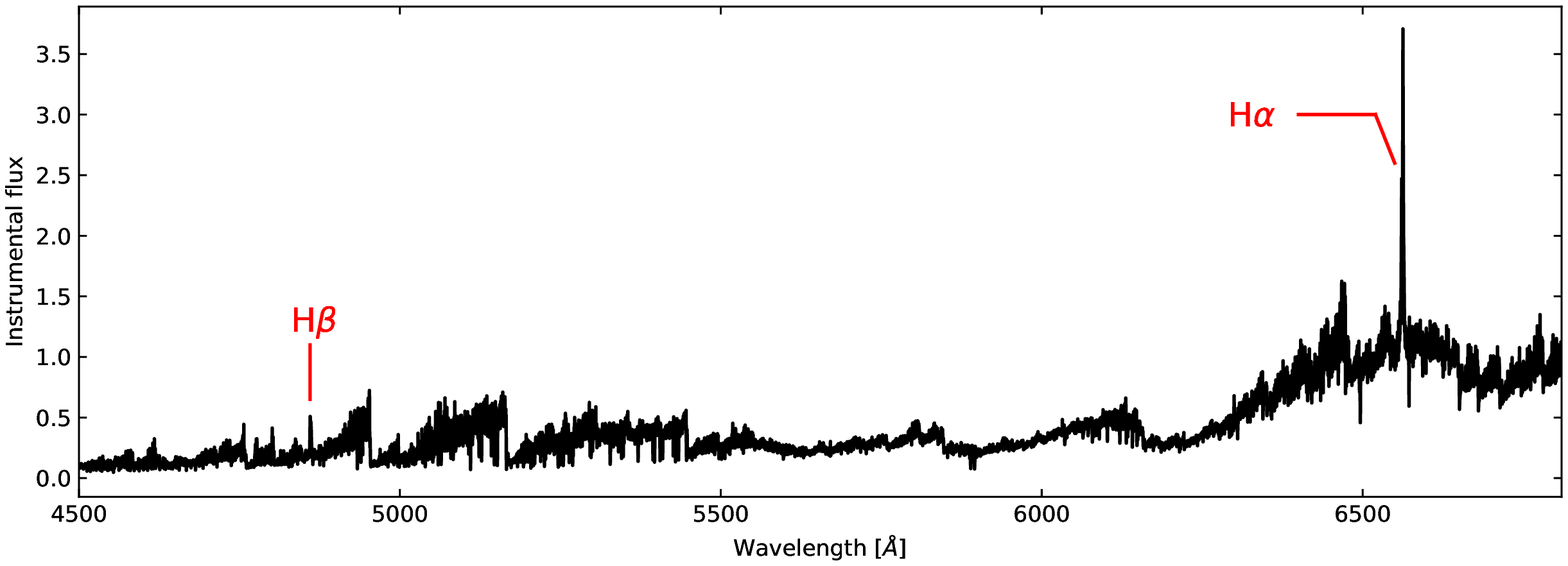}
    \caption{The SOPHIE spectrum of SU~Lyn from the year 2012, when the system was X-ray bright. The spectrum was not flux-calibrated and it is affected by the instrumental flux modulation.}
    \label{fig:SULyn_SOPHIE}
\end{figure*}

We supplemented our observations with the archival spectrum obtained with SOPHIE spectograph mounted on a 1.93m telescope in the Haute Provence Observatory \citep{2004PASP..116..693M,2006tafp.conf..319B}. The SOPHIE spectrum was carried out on 2012-03-12 with an exposure time of 600\,s. A spectrum of 13~Lyr was not subtracted from the SU~Lyn SOPHIE spectrum because in the SOPHIE spectrum the emission lines are significantly stronger than the red giant absorption features (Fig.~\ref{fig:SULyn_SOPHIE}).

We measured spectral lines using two methods. In order to measure equivalent widths (EWs), full-width at half maxima (FWHM) and radial velocities we fitted Gaussian profiles. In order to measure individual pieces of lines consisting of two components (H$\alpha$ and \mbox{[O\,{\sc iii}]}) we fitted two Gaussian profiles. This resulted in an error of the measured EWs of up to 10\%, accuracy of the FWHM of  0.1\,\AA, and the accuracy of the radial velocities of 3~km/s. In the case of the total EWs of lines with two components (H$\alpha$ and \mbox{[O\,{\sc iii}]}) we measured the EWs by integrating spectral region in the $\pm$400~km/s range after the red giant continuum was subtracted. This resulted in an accuracy of 0.1\,\AA. Measurements of emission lines, as well as Na~D1 absorption line, are presented in Tab~\ref{logspec}.

\section{Results}

\subsection{X-ray variability}\label{sec:xray}

SU~Lyn was first observed in the X-ray range by the \textit{ROSAT} satellite in the 1990s (1RXS~J064255.9+552835; \citealt{1999A&A...349..389V}). The optical counterpart of this X-ray source was identified to be SU~Lyn by \citet{2016MNRAS.461L...1M}. In the \textit{Swift} observations of \citet{2016MNRAS.461L...1M} with the energy range of 15--35 keV the system showed a high state between 14th of October 2010 and 1st of August 2012. During the high state, SU~Lyn appeared to maintain a constant, high X-ray luminosity. The remaining \textit{Swift} observations out the X-ray high state showed a constant low X-ray brightness, between the years 2005--2010 and between the years 2012-2015.

\begin{figure}
\resizebox{\hsize}{!}{\includegraphics{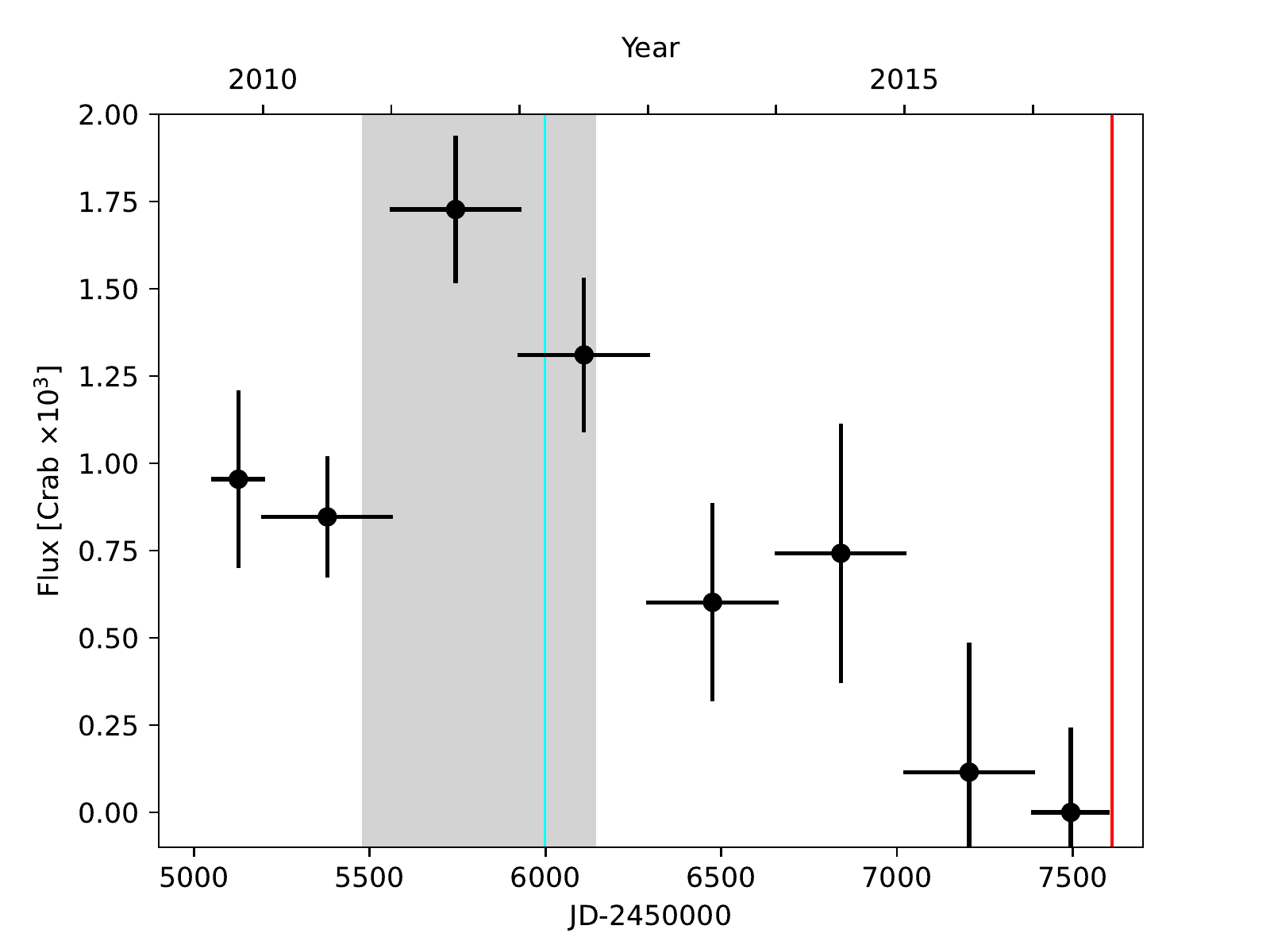}}
    \caption{An X-ray light curve of SU~Lyn in the 4--10~keV range. The red line marks the observation of \citet{2018ApJ...864...46L}. The gray area marks the X-ray high state observed by \citet{2016MNRAS.461L...1M}. The cyan line marks the date of the SOPHIE spectrum.}
    \label{fig:3maxi}
\end{figure}

As opposed to the \textit{Swift} observations, where the system seemed to plateaued at maximum, the data from the MAXI/GSC X-Ray Source Catalog (\textit{3MAXI}; \citealt{2018ApJS..238...32K}) indicates that the system showed a more flare-like behavior (Fig.~\ref{fig:3maxi}). The \textit{3MAXI} observations in the 4--10~keV range are consistent with the results of \citet{2016MNRAS.461L...1M}, where the authors argued that the mass transfer rate decreased after the X-ray high state observed by \citet{2016MNRAS.461L...1M}.   Because of the difference in the energy ranges covered by \textit{3MAXI} and \textit{Swift}  the apparent differences in the light curves can be due to different behavior of SU~Lyn in these two energy ranges. This behavior could be associated with the variable source of absorption in the system \citep{2018ApJ...864...46L}. However, the X-ray spectrum seems to consist of only one component \citep{2016MNRAS.461L...1M,2018ApJ...864...46L}. Therefore, it is possible that the \textit{3MAXI} observations have a higher signal to noise ratio and the flare-like behavior reflects the real variability. This implies the X-ray maximum occurred in the year 2011 rather than between the years 2010 and 2012. Hence, we will refer to the X-ray maximum as an X-ray flare rather than the bright state that was used in the literature. There are no X-ray observations carried out after 2016 that are published, therefore it is not certain whether the X-ray luminosity continued to decrease. However, given the spectroscopic variability (Section~\ref{sec:spec}) it seems likely that no X-ray flare occurred after the 2010--2012 brightening.

\subsection{Photometric variability}\label{sec:phot}

The historical light curve in the of SU~Lyn in the m$_\mathrm{pg}$ and $B$ filters did not show any significant changes in the visual brightness during the last century (Fig.~\ref{fig:lc_hist}). While there are significant gaps in the data, this suggests that there were no nova-like or Z-And type outbursts in the recent history of the system.

Observations just before the X-ray flare show SU~Lyn at $V\simeq$~8.5~mag (Fig.~\ref{fig:lc_hist}). WASP observations, which were carried out with a photometric filter relatively close to the $V$ band \citep{2006PASP..118.1407P}, suggest that there were no major changes in the SU~Lyn brightness up to a few years before the X-ray flare. During the high state, the system faded by $\sim 0.5$~mag in the $V$ band. Conversely, after the X-ray flare, the system experienced a brightening by $\sim 0.5$~mag in the $V$ band. In 2014 the brightness of SU~Lyn returned to $V\simeq$~8.5~mag. After that, SU~Lyn gradually increased its visual brightness and reached $V\simeq$~8.0~mag between the years 2014 and 2020. \citet{2021arXiv210402686M} showed that the $B-V$ color did not change significantly during this increase in the visual brightness.

The visual variability does not follow the UV observations, where the maximum flux was observed in 2015  \citep{2016MNRAS.461L...1M,2018ApJ...864...46L,2021MNRAS.500L..12K}. However, no UV observations exist from the time of X-ray flare and 2015 observations are the first observations after the flare. Only two UV observations from the years 2006 and 2007 exist from before the X-ray flare \citep{2021MNRAS.500L..12K}. A similar decrease in the UV component flux was observed between 2015 and 2017 in the optical spectra of \citet{2021arXiv210402686M}.

The amplitude of variability in the $R$ band was significantly higher compared to the $V$ band. SU~Lyn brightness changed from $R\simeq$8.3~mag to $R\simeq$6.9~mag (Fig.~\ref{fig:lc_hist}). The $R$ brightness from before the X-ray flare is less precisely estimated due to the scarcity of observations.  A brightness of $R$=7.60~mag reported by \citep{2003AJ....125..984M} may suggest that variability in the $R$ band follows the variability in the $V$ band, but with a higher amplitude.

SU~Lyn semi-regular pulsations were observed with a period of 126~days between the years 1932 and 1938 \citep{1955AN....282...73K}. In the ASAS-SN $g$ data we find semi-regular pulsations with a period of $\sim$134~days. A similar variability was observed by \citet{2021arXiv210402686M}, where the authors noted that they did not find any strict periodicity based on data covering a similar time period. We note that our results are consistent with that of \citet{2021arXiv210402686M}, since we do not claim a strict periodicity, but rather a quasi-period that is expected to vary in semi-regular pulsators. In order to study the variability of the pulsation period, we analyzed the historical  m$_\mathrm{pg}$ observations. We find that the pulsations were strongest between the years 1910 and 1920 with a period of $\sim$118~days. The pulsations seem weaker before and after that dates, including the period observed by \citep{1955AN....282...73K}. However, we note that the historical data are not of sufficient quality for conclusive results. Interestingly, the pulsations were not present in data from the DIRBE infrared survey \citep{2010ApJS..190..203P} taken between the years 1990 and 1993. This does not imply that the pulsations were not present at this time, as similar semi-regular pulsations are not always seen in the DIBRE data. For example, while semiregular-pulsators UU~Dra and U~Del were classified as variable objects by DIBRE, no variability was detected from semiregular-pulsators  TT~Peg and SV~Peg \citep{2010ApJS..190..203P}. In summary, all the available data can suggest that there is a possible evolution of pulsations in SU~Lyn, but the results are inconclusive. 


\subsection{Spectral variability}\label{sec:spec}

The first spectra of SU~Lyn were carried out by \citet{1972MNRAS.158...23F}, where no emission lines were noted. However, they did not present the spectrum itself. \citet{2016MNRAS.461L...1M} showed that weak emission lines were present in SU~Lyn spectrum. This was later confirmed by \citet{2021MNRAS.500L..12K}, which also showed that there are emission lines present in the UV spectral range. Because only weak emission lines were present in the optical spectrum of SU~Lyn, \citet{2016MNRAS.461L...1M} suggested that SU~Lyn is a hidden symbiotic star. The limitation of the \citet{2016MNRAS.461L...1M} hypothesis is that the only available spectra analyzed by them were from a limited period of time. In particular, the earliest spectrum of \citet{2016MNRAS.461L...1M} was from the year 2016, four years after the X-ray flare.

\subsubsection{Balmer lines}

\begin{figure}
\resizebox{\hsize}{!}{\includegraphics{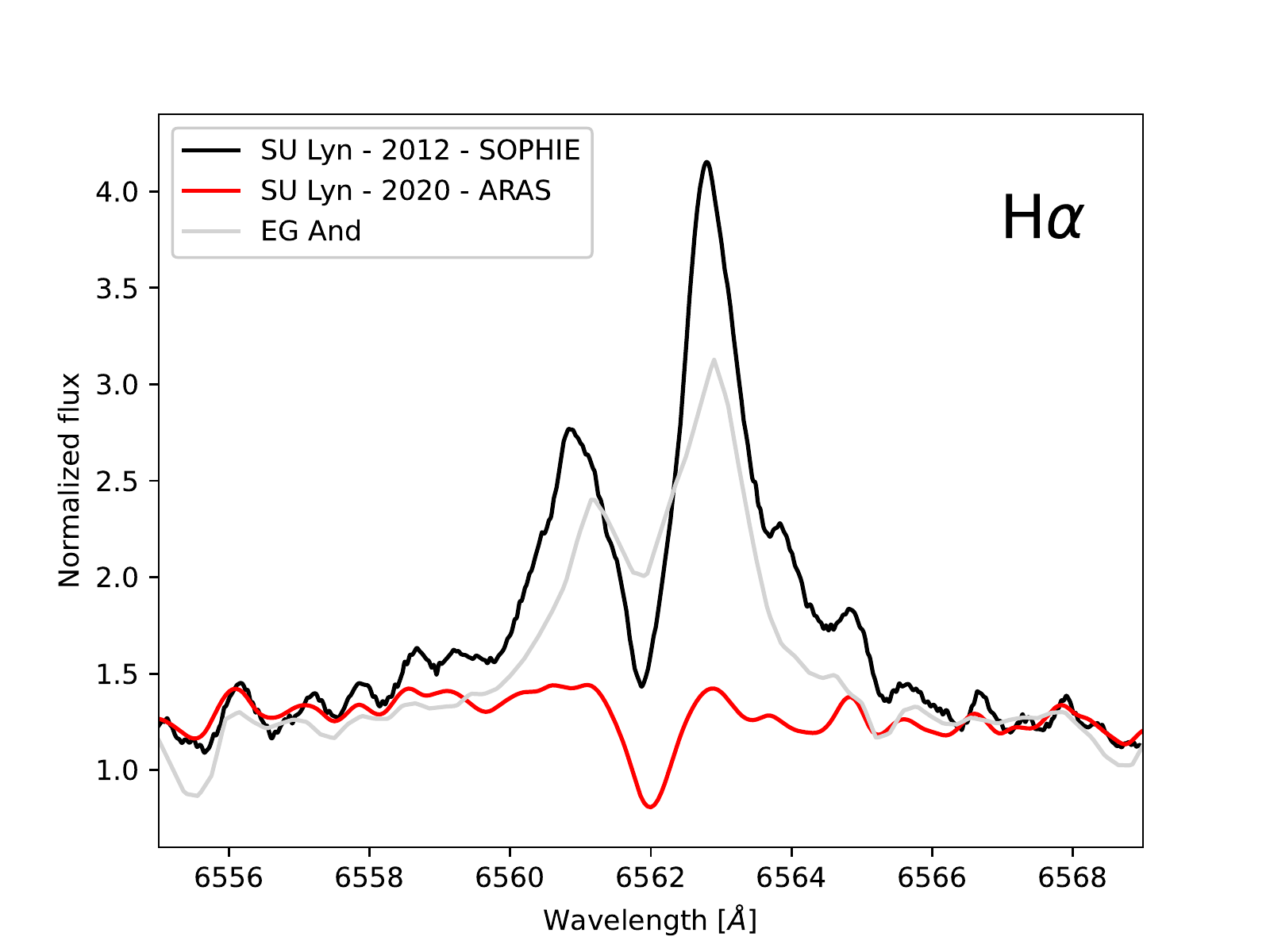}}
\resizebox{\hsize}{!}{\includegraphics{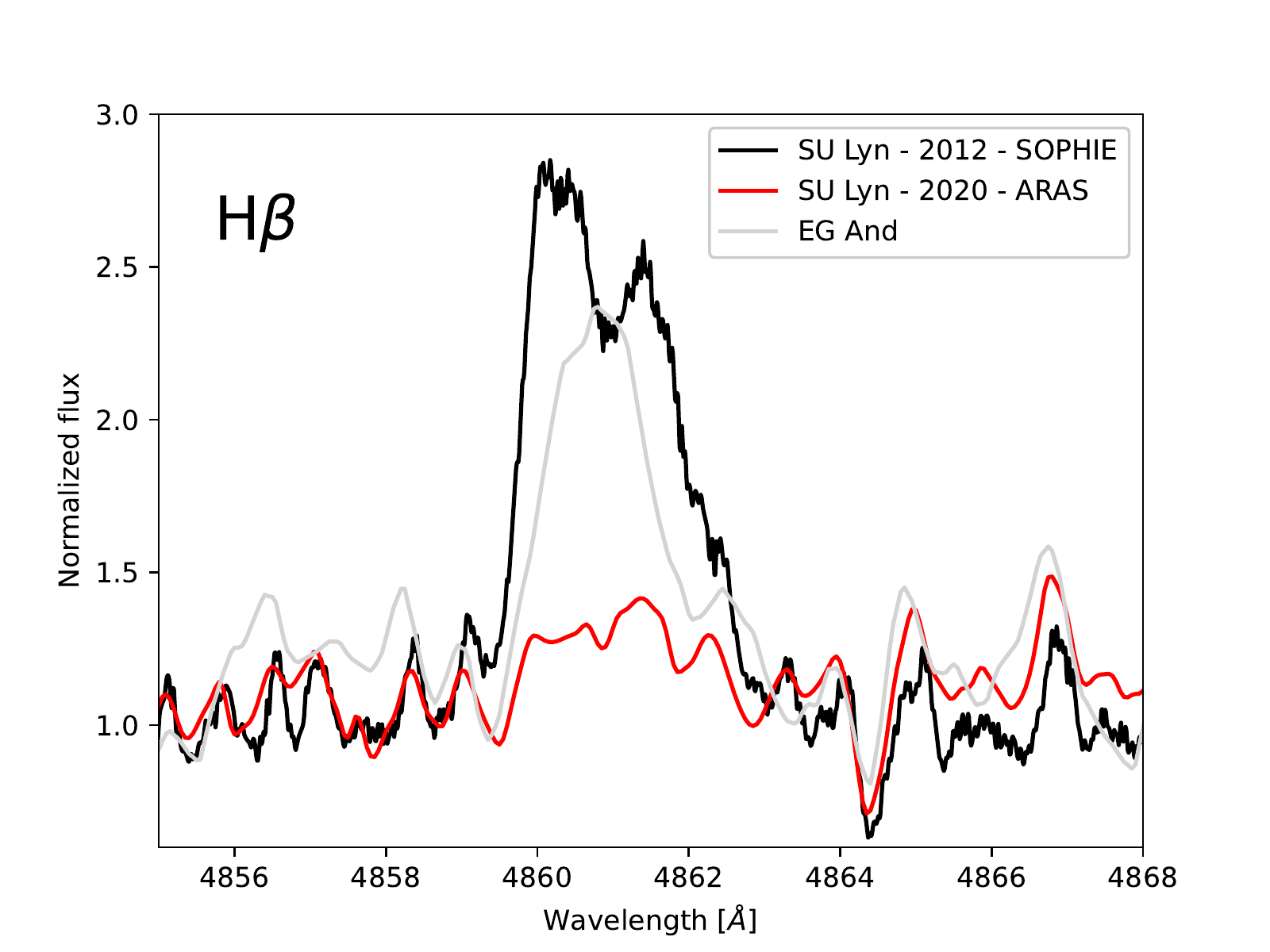}}

\caption{Balmer lines in SU~Lyn from the time of X-ray high state (2012) and out of the X-ray high state (2020). A classical symbiotic star spectrum of EG~And is shown for comparison. The EG~And spectrum was carried out by Fran\c cois Teyssier on 26.01.2017 and was taken from the ARAS database. The date of EG~And spectrum was chosen to be close to the median H$\alpha$ emission strength observed in this object by \citet{2021A&A...646A.116S}. The EG~And spectrum has been shifted to fit the SU~Lyn radial velocity.}
    \label{fig:balmer_SOPHIE}
\end{figure}

The SOPHIE spectrum is the only spectrum available from the time near the X-ray flare (Fig.~\ref{fig:lc_hist}). In the SOPHIE spectrum, strong Balmer lines are present, in contrast to the spectrum presented by \citet{2016MNRAS.461L...1M}. While systematic comparison of these lines with the entire population of known symbiotic stars is beyond the scope of this work, a comparison with EG~And shows that these lines were as strong as at lest in some of the other symbiotic stars (Fig.~\ref{fig:balmer_SOPHIE}), while they were significantly fainter afterwards \citep{2016MNRAS.461L...1M}.

\begin{figure}
\resizebox{\hsize}{!}{\includegraphics{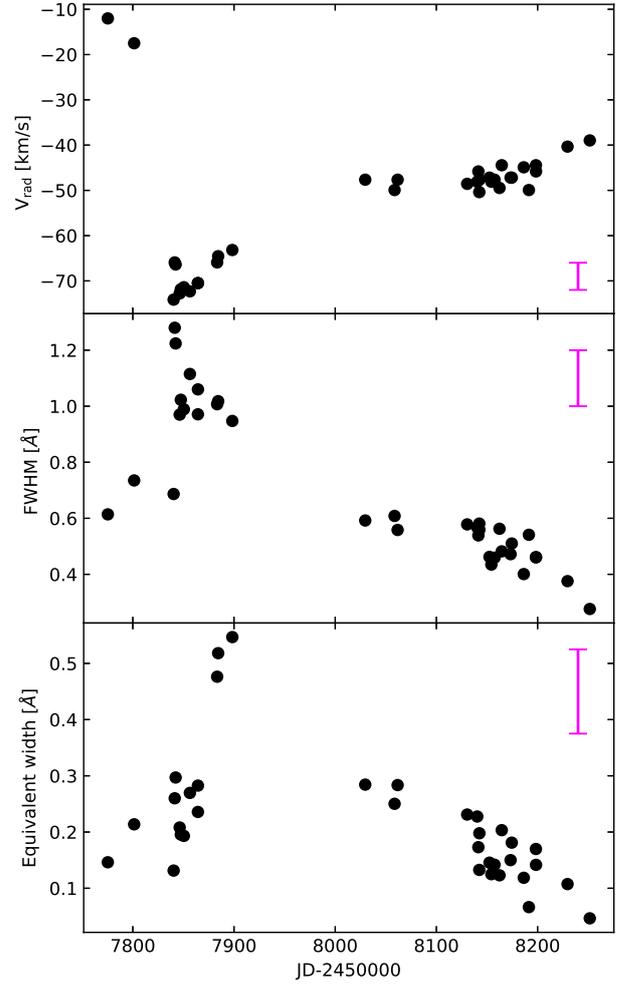}}
\caption{Evolution of an absorption component of the P~Cygni profile in the H$\alpha$ line. The magenta error bars indicate the typical error of the measurements.}
    \label{fig:Halpha_measured}
\end{figure}

Interestingly, no forbidden lines such as \mbox{[O\,{\sc iii}]} are present in the SOPHIE spectrum. The Balmer and \mbox{[O\,{\sc iii}]} lines are present on all of the ARAS spectra. Our spectral range does not cover the  \mbox{[Ne\,{\sc iii}]}~3868 line reported by \citet{2016MNRAS.461L...1M} and \citet{2021MNRAS.500L..12K}. On the first ARAS spectrum from JD=2457513, the H$\alpha$ line has a profile typical for a symbiotic star, with a broad emission (FWHM=2.5\,\AA) and a narrow absorption from the red giant. After removing the red giant features, which remained constant throughout our observations, the second ARAS spectrum from JD=2457775 revealed a P~Cygni profile. The absorption part of the P~Cygni profile evolved with time, changing its width, the mean radial velocity, as well as depth (Fig.~\ref{fig:all_spec}). We measured its evolution by fitting two Gauss functions, one for the evolving absorption and one for the emission part of the P~Cygni profile. The radial velocity of the absorption first increased sharply towards bluer wavelengths and then decreased the velocity gradually (Fig.~\ref{fig:Halpha_measured}). Similarly, the width of the absorption first rose sharply and then decreased gradually. Both the radial velocity and width of the P~Cygni absorption reached their maxima at JD$\simeq$2457840. While the equivalent width of the absorption experienced behavior similar to the other two studied parameters, its maximum was delayed by at least $\sim$50~days (Fig.~\ref{fig:Halpha_measured}). The reason for the delay is unclear, but the delay may be due to the pulsations of the red giant, which would change the flux at the base of the line and introduce a shift in the measured equivalent width. The last spectrum in which the P~Cygni profile is visible is from JD=2458252 and on all further spectra only the emission is present. It seems that the P~Cygni profile observed by us is associated with a ,,second absorption component'' discovered by \citet{2021arXiv210402686M}. However, \citet{2021arXiv210402686M} reported on the emergence of this absorption profile only at the time of our observed absorption radial velocity maximum, while we observed a gradual increase in the radial velocity. Moreover, \citet{2021arXiv210402686M} reported that the new absorption component merged with the red giant absorption line, while we observed a gradual decrease in its equivalent width.

\begin{figure}
\resizebox{\hsize}{!}{\includegraphics{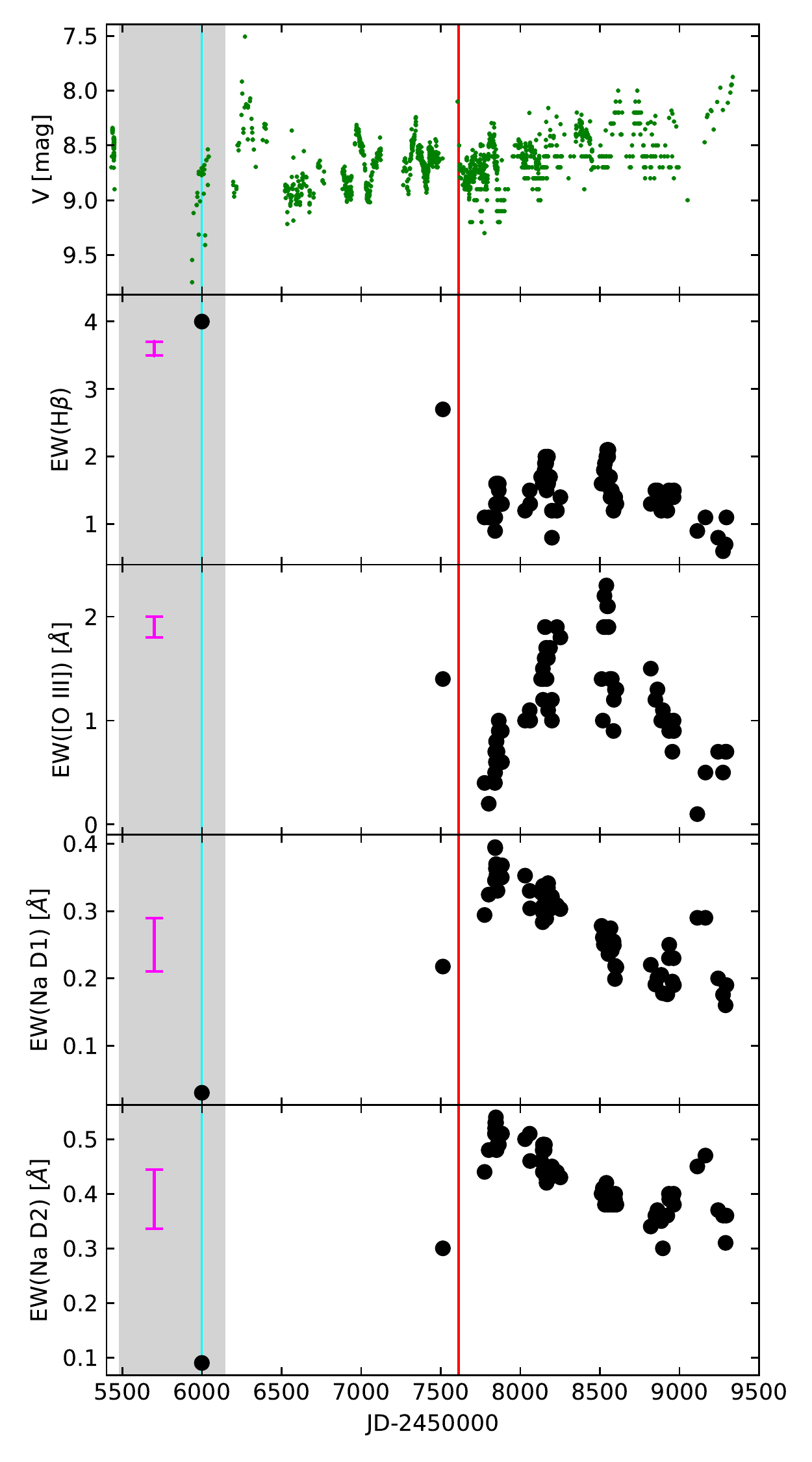}}
\caption{Evolution of the equivalent widths of emission lines and blue-shifted component on Na~D absorption lines in SU~Lyn compared to the photometric variability in the $V$ band. All emission lines were measured after subtraction of a reference spectrum of 13~Lyr. The measurements of \mbox{[O\,{\sc iii}]} emission line equivalent width included simultaneous measurement of the two line components.  The magenta error bars indicate the typical error of the measurements.}
    \label{fig:SULyn_all}
\end{figure}

The equivalent width of all of the lines showed significant variability (Fig.~\ref{fig:SULyn_all}). For the Balmer lines analysis, we restricted ourselves to analyzing the H$\beta$ line since it did not seem to show the P~Cygni profile. However, we note that the emission component of H$\alpha$ line seemed to follow the same variability. The Balmer lines were strongest in the SOPHIE spectrum, right after the X-ray flare. On the next available spectra, equivalent widths of H$\beta$ were decreasing, until they reached a minimum at JD$\simeq$2457800. Afterward, the equivalent width of H$\beta$ was increasing, until it reached a local maximum at JD$\simeq$2458550. After the local maximum, the H$\beta$ strength continued to decrease until our last observation. This result is consistent with the observations of \citet{2021arXiv210402686M}, where the H$\alpha$ emission seemed to reach a local maximum in the year 2018. It seems that no photometric variability accompanied the changes in H$\beta$ emission line. We note, that the H$\beta$ equivalent widths showed variability on shorter timescales than the one discussed here. However, this variability can be attributed to pulsations of the red giant and resulting changes in the continuum level. These changes would not be removed by subtracting the spectrum of 13~Lyr, as both spectra of SU~Lyn and 13~Lyr were normalized to the local continuum before the subtraction.

\subsubsection{\mbox{[O\,{\sc iii}]}~5007 line}\label{sec:oiii}

\begin{figure}
\resizebox{\hsize}{!}{\includegraphics{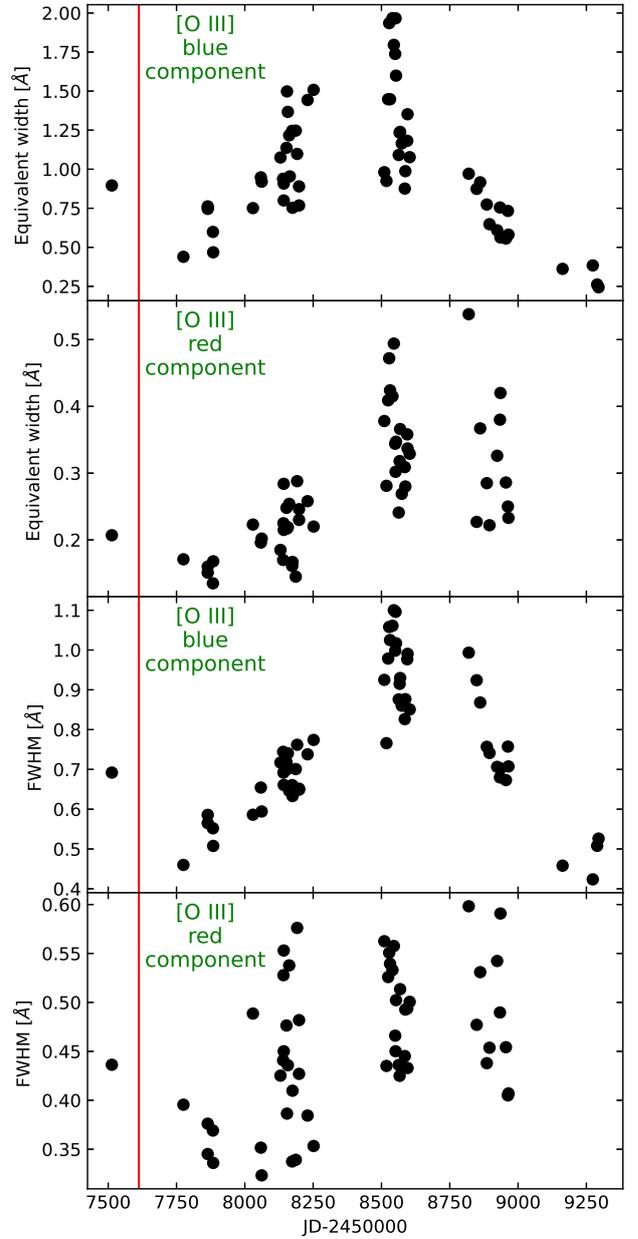}}
\caption{Variability of the blue and red components of the \mbox{[O\,{\sc iii}]}  emission line.}
    \label{fig:OIII_measured}
\end{figure}

\begin{figure*}
	\includegraphics[width=\textwidth]{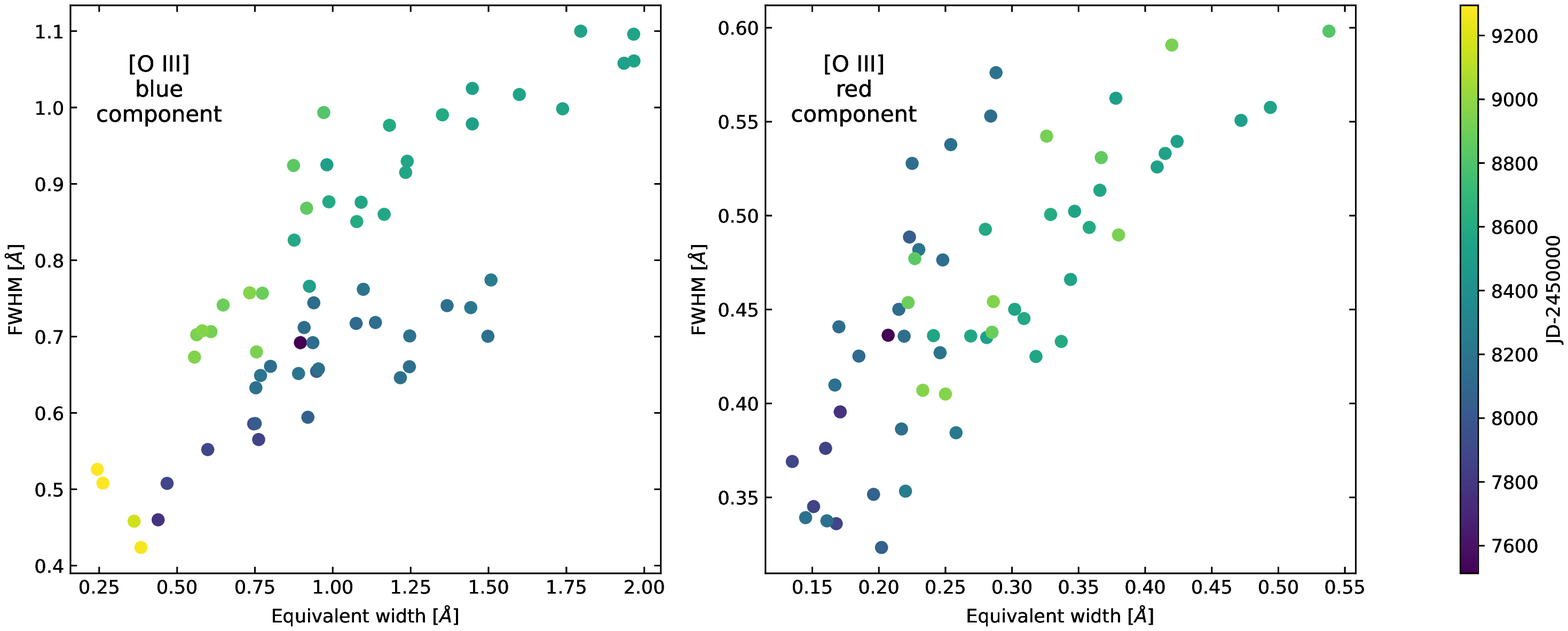}
    \caption{Relation between the line width and line strength of two \mbox{[O\,{\sc iii}]}  emission line components -- the blue component (left panel) and red component (right panel).}
    \label{fig:OIII_components}
\end{figure*}

The integrated \mbox{[O\,{\sc iii}]}  emission line flux showed variability similar to the H$\beta$ line (Fig.~\ref{fig:SULyn_all}). However, upon closer inspection it is clear that sometimes the \mbox{[O\,{\sc iii}]} emission consist of two components (Fig.~\ref{fig:all_spec}). The blue component was stronger and was located at -12~km/s relative to the red giant radial velocity. The red component of the \mbox{[O\,{\sc iii}]}  emission was located at +39~km/s  relative to the red giant radial velocity. This stands in contrast to the H$\beta$ emission, which seemed to be red-shifted on average, but remained consistent with the red giant radial velocity within our error of 3~km/s. Both the components seemed to follow the variability of the H$\beta$ line strength (Fig.~\ref{fig:OIII_measured}). The red component was not detected in our last observations. Both components line widths seemed correlated with the line strengths (Fig.~\ref{fig:OIII_components}). However, the blue component of \mbox{[O\,{\sc iii}]} line seemed to be narrower than during the rise of the line strength than during the decrease in the line strength. The opposite might be true for the red component, but the data accuracy for this component of the \mbox{[O\,{\sc iii}]} line is too low for a conclusive result.

\subsubsection{Na~D lines}

In the SU~Lyn spectra Na~D absorption lines are present at the red giant radial velocity. These lines are expected for the spectral type of the red giant in SU~Lyn and are also present in the spectrum of 13~Lyr. Moreover, the stellar Na~D lines are blended with an interstellar component detected at -4.1~km/s by \citet{2016MNRAS.461L...1M}. However, an additional blue-shifted component of the Na~D lines is present at the radial velocity of -88~km/s (Fig.~\ref{fig:all_spec}). While the Na~D lines connected to the red giant remained constant, the blue-shifted absorption showed strong variability. Namely, the blue-shifted Na~D absorption showed a maximum at JD$\simeq$2457800, when the emission lines showed a minimum (Fig.~\ref{fig:SULyn_all}). After that date, the blue-shifted Na~D absorption was decreasing independently from the variability of emission lines. The exception was a local maximum of the blue-shifted absorption at JD$\simeq$2459112, where it seemed that the \mbox{[O\,{\sc iii}]} emission experienced a local minimum (Fig.~\ref{fig:all_spec})

The changing equivalent width of the blue-shifted Na~D absorption can be connected to a changing circumstellar component of extinction. This is consistent with the observed changing column density towards the white dwarf observed by \citet{2018ApJ...864...46L}. The equivalent width of  Na~D1 line can be used to calculate the change in reddening \citep{1997A&A...318..269M}. The minimum Na~D1 equivalent width measured at JD=2456000 corresponds to circumstellar extinction of E($B-V$)=0.01~mag. Afterward, the change in reddening is following the changes in equivalent widths of Na~D lines, with the maximum circumstellar extinction at E($B-V$)=0.18~mag at JD=2457841. The circumstellar extinction reached a significantly larger value than the constant interstellar extinction E($B-V$)=0.07~mag estimated by \citet{2016MNRAS.461L...1M} on JD=2457409  We note that the \mbox{[K\,{\sc i}]}~7699 line that should behave similarly to Na~D lines was outside the spectral range of our spectra.

\subsection{Red giant radial velocity}\label{rvsearch}

We searched for possible changes in the red giant radial velocity using cross-correlation of SU~Lyn spectra with 13~Lyr spectrum. We did not find any significant variability within our accuracy of 3~km/s. This is consistent with the reported limit on radial velocity variability of  $<2$~km/s reported by \citet{2021arXiv210402686M}. The mean red giant radial velocity in our spectra was -22~km/s.

\subsection{The red giant evolution status}\label{sec:evol}

\citet{2016MNRAS.461L...1M} estimated a distance to SU~Lyn of 640$\pm$100~pc. A similar distance of 659$^{+46}_{-40}$~pc was obtained using \textit{Gaia} DR2 data  \citep{2018AJ....156...58B} and a distance of 710$^{+31}_{-35}$~pc using \textit{Gaia} DR3 data \citep{2021AJ....161..147B}. The effective temperature of the RG estimated in \textit{Gaia} DR2 was $T_{\mathrm{eff,RG}}$=3544$^{+249}_{-256}$~K, while the luminosity $L_{\mathrm{RG}}$=2261$\pm$176~L$_\odot$ \citep{2018A&A...616A...1G}. No red giant parameters were estimated in \textit{Gaia} DR3 \citep{2021A&A...649A...1G}. We note that the distance and luminosity estimated by \textit{Gaia} may be unreliable since currently the  \textit{Gaia} archive does not account for the astrometric motion due to the binary nature of the system \citep[e.g.][]{2011A&A...536A..55B}. The red giant temperature of $T_{\mathrm{eff,RG}}\simeq$3565~K was estimated by \citet{2019ApJS..240...21A}. We estimated the effective temperature and luminosity of the RG by fitting models to the spectral energy distribution (SED) using the VO Sed Analyzer \citep[VOSA;][]{2008A&A...492..277B}. 

For the SED fit we used data from TYCHO-2 \citep{2000A&A...355L..27H}, Pan-STARRS \citep{2016arXiv161205243F}, 2MASS \citep{2006AJ....131.1163S}, AKARI/IRC \citep{2010A&A...514A...1I}, IRAS PSC/FSC \citep{2015A&C....10...99A} and unWISE \citep{2019ApJS..240...30S} surveys. In order to avoid contamination from the white dwarf and any dust emission we limited the fit only to data ranging from 420 to 52000~nm. Since all of the employed observations were obtained before the X-ray flare and the changes in the circumstellar reddening, we limited the reddening correction to only the interstellar component equal to E($B-V$)=0.07~mag \citep{2016MNRAS.461L...1M}. We also assumed the distance of 710$^{+31}_{-35}$~pc \citep{2021AJ....161..147B}.

\begin{figure}
\resizebox{\hsize}{!}{\includegraphics{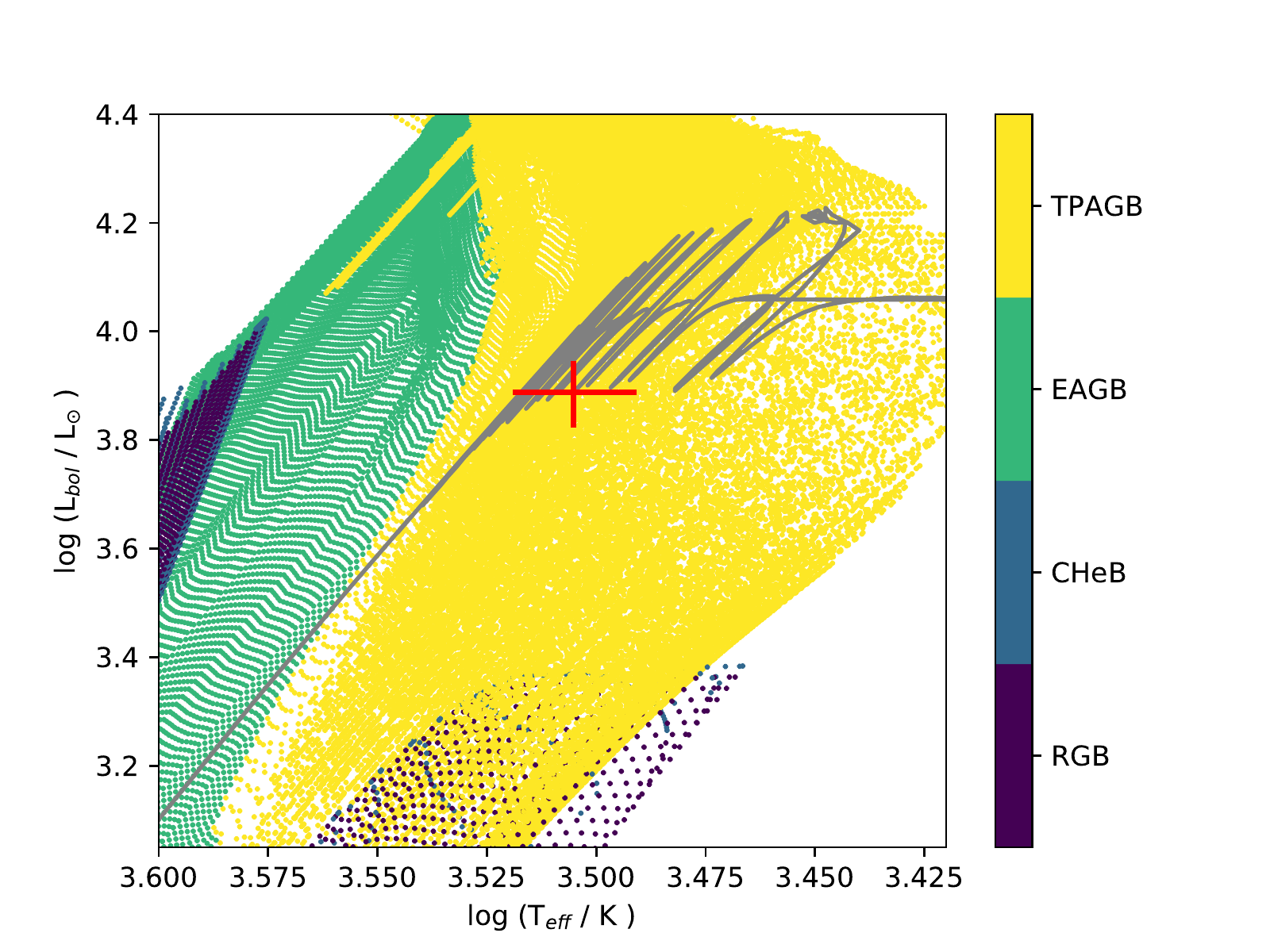}}
\caption{The stellar parameters of RG in SU~Lyn (Table~\ref{tab:param}; red cross) compared to single star evolution models from MIST. The tracks have been computed using zero-age main sequence masses in the range of 0.5-8~M$_\odot$ with a step of 0.05~M$_\odot$. The evolutionary track best fitting VOSA estimates is highlighted with a gray line.}
    \label{fig:SULyn_HR}
\end{figure}

As a result of the SED fit, we obtained T$_{\mathrm{eff,RG}}$=3200$\pm$100~K and L$_{\mathrm{RG}}$=7700$\pm$1000~L$_\odot$. The estimated luminosity was higher than the one observed by \textit{Gaia}, while the effective temperature was lower. We note that our estimate of the red giant effective temperature is consistent with the M5 spectral type derived by \citet{2016MNRAS.461L...1M}, while the \textit{Gaia} effective temperature is too high for this spectral type. Hence, the parameters estimated using VOSA are more consistent. In order to estimate the evolutionary status of the RG, we compared the red giant parameters to the MESA Isochrones and Stellar Tracks (MIST; \citealt{2016ApJS..222....8D}, \citealt{2016ApJ...823..102C}). Using T$_{\mathrm{eff,RG}}$ and L$_{\mathrm{RG}}$ from \textit{Gaia} and VOSA we identify the donor in SU~Lyn as a relatively young thermally pulsating RG (Fig.~\ref{fig:SULyn_HR}).  Using the parameters derived with VOSA we estimate log(Age/yr)=8.70$^{+0.36}_{-0.23}$ and M$_{\mathrm{RG}}$=2.78$^{+0.62}_{-0.64}$~M$_\odot$.  The RG stellar parameters are summarized in Table~\ref{tab:param}. If the red giant in SU~Lyn experienced a thermal pulse in recent history, this can be associated with changes in the pulsation period, which would be consistent with discussion in Section~\ref{sec:phot}. Changes in the observed pulsation period due to a thermal pulse were e.g. observed in another semiregularly pulsating star RU~Vul \citep{2016AN....337..293U}.

%

\begin{table}
 \centering
  \caption{ Summary of the derived stellar parameters of the RG in SU~Lyn. }\label{tab:param}
  \begin{tabular}{|ccccccccc|}
  \hline
T$_{\mathrm{eff,RG}}$ [K]	&	L$_{\mathrm{RG}}$ [L$_\odot$] & M$_{\mathrm{RG}}$ [M$_\odot$] & R$_{\mathrm{RG}}$ [R$_\odot$] & log(Age/yr)	\\
\hline																	
	3200$\pm$100		&	7700$\pm$1000	&	2.78$^{+0.62}_{-0.64}$	& 280$\pm$40	& 8.70$^{+0.36}_{-0.23}$	\\		

\hline
\end{tabular}
\end{table}

The red giant 2MASS magnitudes $J$=2.979 and $K$=1.618~mag translate to a Weisenheit index equal to $W_{JK}$=0.684~mag. Scaling this index to the Large Magellanic Cloud distance of 49.59~kpc \citep{2019Natur.567..200P} we obtain $W_{JK}^{\mathrm{LMC}}$=10.15~mag. Together with the pulsation period of $\mathrm{P}_{\mathrm{pul}}$=126~days, this indicates that the red giant in SU~Lyn belongs to a C' sequence \citep{2015MNRAS.448.3829W}. Using our VOSA analysis we estimated the red giant radius to be R$_{\mathrm{RG}}$=280$\pm$40~R$_\odot$. This gives a pulsation constant $Q=\mathrm{P}_{\mathrm{pul}}(\mathrm{M}_{\mathrm{RG}}/ \mathrm{M}_\odot)^{1/2}(\mathrm{R}_{\mathrm{RG}}/ \mathrm{R}_\odot)^{-3/2}\simeq$0.044~days. This pulsation constant is consistent with pulsations in the first overtone \citep[e.g.][]{1982ApJ...259..198F}.

\section{Discussion}

SU~Lyn shows a very complex behaviour. In particular, optical emission line changes correlated with photometric and X-ray variability. Whereas it might not be easy to identify mechanism(s) behind the variability,  it is clear that not only the X-ray emission showed a transient brightening, but also the optical appearance of the system was transient in nature.

\subsection{Nature of variability}
 
SU~Lyn is the first object of its kind ever observed \citep{2016MNRAS.461L...1M}. Because of this, a comparison to variability of other binary systems is not trivial. However, we would like to propose a few directions which future studies could explore. While not much is known about the system orbital parameters, the only certain thing is that the system consists of a white dwarf and a red giant \citep{2016MNRAS.461L...1M,2018ApJ...864...46L,2021MNRAS.500L..12K}. Therefore, all proposed scenarios will include both binary components. Moreover, SU~Lyn was observed as an X-ray emitting object before the X-ray flare \citep{2016MNRAS.461L...1M}, therefore a possibility of recurring or periodic X-ray flashes will be explored.

The simplest scenario involves a highly eccentric orbit. During a periastron passage, the tidal forces would lead to an enhanced mass loss from the red giant, allowing for an increased mass transfer rate. This scenario could be confirmed if the discussed SU~Lyn behaviour would be repeatable. However, the orbital period of SU~Lyn is unknown and the object should be monitored continuously for at least two additional cycles to confirm this hypothesis. A similar, quasi-periodic enhanced mass transfer rate near a periastron passage was observed in a symbiotic stars BX~Mon \citep{2012BaltA..21..172A,2013AcA....63..405G}, CD-43~14304 \citep{2013AcA....63..405G}, and  GX~1+4 \citep{2017A&A...601A.105I}.

The mechanism triggering the variability could be related to a potential He-shell flash of the red giant. Such a flash could be related to a mass ejection event  \citep[e.g.][]{2007A&A...470..339M}, which would lead to a temporary increase in the mass accretion rate by the white dwarf. Furthermore, it is possible that the white dwarf would then heat up and be able to ionize the material close to the red giant, potentially ionizing the wind acceleration zones and decreasing the radiation pressure. The material would build up in these wind acceleration zones until it would cool and the radiation pressure could build up again, ejecting a shell of matter. When this shell would encounter the white dwarf, it would be partially accreted. This cycle of heating material in the wind acceleration would then repeat, leading to the observed variability of SU~Lyn. 


Independently from the mechanism triggering the variability, the variability is associated with a strong mass loss from the system. This is demonstrated by the detection of a transient P~Cygni profile and appearance of Na~D lines at a high blue-shift. Mass loss in symbiotic stars can lead to a binary-induced spiral structure around the binary \citep[e.g.][]{2012BaltA..21...88M,2016MNRAS.457..822B}. Presence of such a spiral could explain the variability of the [O\,{\sc iii}]~5007 line (Section \ref{sec:oiii}). Namely, the presence of a blue and red components of [O\,{\sc iii}] line could originate in two opposite spiral arms. The apparent increase of width of the blue component of the [O\,{\sc iii}] line could be explained if a new, closer loop of the spiral formed later in time. This more distant loop would have a lower radial velocity due to interaction with the interstellar material. However, the difference  would not be larger and both blue components would blend in our observations, resulting in a single, wider line. A similar structure is present in a nearby symbiotic star, R~Aqr, where the spiral structure was formed due to a highly eccentric orbit \citep{2021A&A...651A...4B}. The circumstellar material present in R~Aqr shows radial velocity amplitude of $\sim$10~km/s \citep{2021A&A...651A...4B}, which is of the same order of magnitude as the observed radial velocities of the blue and red components of the [O\,{\sc iii}] line in SU~Lyn. The caveat is that the spiral structure in R~Aqr consists of dusty molecular dust, while the material around SU~Lyn is ionized. This difference could be due to the fact that the structure around SU~Lyn is newly formed and transient in nature, implying shocks in the outer region of the forming spiral ionizing the inner parts of the loop. Alternatively, the white dwarf in SU~Lyn could be more efficient in ionizing the material in the spiral structure.


\subsection{SU~Lyn as a transient symbiotic star}

Symbiotic stars are defined as interacting binaries with a red giant donor and a compact companion. An observational manifestation of this interaction is the presence of emission lines in the optical part of the spectrum. Because no strong emission lines were present in the optical part of the SU~Lyn spectrum and there was evidence of high mass transfer rate during the X-ray flare, \citet{2016MNRAS.461L...1M} argued that SU~Lyn is a hidden symbiotic star. This is because a high mass accretion rate indicates interaction between the red giant and the white dwarf, making the system an interacting binary system -- a symbiotic star. However, we showed that when the mass transfer rate was high, the optical emission lines were as strong as at least in some other classical symbiotic stars (Fig.~\ref{fig:balmer_SOPHIE}). Therefore, when the system was interacting, the spectrum of SU~Lyn was that of a typical symbiotic star, and the nature of the system was not ''hidden'' in the optical spectra.

The problem with the classification of SU~Lyn as a symbiotic star is due to the intensity of the binary interaction. The mass accretion rate changed from  $\sim$1.0$\times$10$^{-8}$~M$_\odot$\,yr$^{-1}$ during the X-ray flare \citep{2016MNRAS.461L...1M} to 1.5--2.2$\times$10$^{-10}$~M$_\odot$\,yr$^{-1}$ when system was X-ray faint \citep{2018ApJ...864...46L}. The mass accretion rate outside of the X-ray flare is smaller than the mass accretion rate of  $\sim$10$^{-9}$~M$_\odot$\,yr$^{-1}$ that is the minimum mass accretion rate typical for a symbiotic star \citep{1984ApJ...279..252K}.

The population of symbiotic stars is studied assuming that as a result of interaction between the system components the mass accretion rate is high enough to power the production of emission lines. In theoretical studies, this is quantified by assuming that the accretion powered luminosity is higher than at least 10~$L_\odot$ \citep[e.g.][]{1992MNRAS.256..177M,2006MNRAS.372.1389L,2020MNRAS.496.3436B}. The white dwarf mass in SU~Lyn was estimated to be in the range of 0.69--0.92~M$_\odot$ \citep{2018ApJ...864...46L}. Using a white dwarf radius calculated with eq. 15 of \citet{1988ApJ...332..193V} the accretion power luminosity of the white dwarf changed from 18--33~L$_\odot$ during the X-ray flare to <0.5~L$_\odot$ during \citet{2018ApJ...864...46L} observations. Therefore, the accretion powered luminosity was high enough to be classified as a symbiotic star during the X-ray flare and not high enough after the flare. Again, this is confirmed by the presence of emission lines with strengths typical for a symbiotic star during the X-ray flare and their absence afterward (Fig.~\ref{fig:balmer_SOPHIE}).

In conclusion, all metrics used to classify a system as a symbiotic star -- the presence of emission lines, mass accretion rate, and accretion powered luminosity -- point at that SU~Lyn could be classified as a symbiotic star only during a short period. Therefore, SU~Lyn can transition between being a hidden and a classical symbiotic star.  \citet{2016MNRAS.461L...1M} argued that the known population of symbiotic stars is biased in favor of shell-burning systems and SU~Lyn is a hidden accretion powered system. This could be interpreted as that SU~Lyn would be a classical symbiotic star if there was shell-burning present in the system. However, the presence of shell burning requires mass accretion rate between $\sim$10$^{-7}$ and $\sim$$10^{-8}$~M$_\odot$\,yr$^{-1}$ depending on the white dwarf mass  \citep[e.g.][]{2007ApJ...663.1269N}. Therefore, SU~Lyn could never be a shell-burning system due to the low mass transfer rate -- the system is not interacting strongly enough. However, if the white dwarf is not a newly formed remnant after the asymptotic giant branch phase of evelution, the nuclear-burning shell must be ignited. Therefore, all symbiotic stars were first accreting only symbiotic stars, and experience at least one thermonuclear nova eruption. A similar transition was recently observed  in a symbiotic star AG~Peg \citep[e.g.][]{2016MNRAS.461.3599R,2016MNRAS.462.4435T,2017A&A...604A..48S}.  Hence,  whether or not the SySt can sustain the shell-burning after ignition depends mainly on how strong the system components are interacting.

Another object that showed variability similar to SU~Lyn is DASCH~J075731.1+201735. This object showed a Z~And type outburst in the 1940s, classifying it as a shell-burning symbiotic star at that time. However, it now appears as a non-interacting binary with a red giant and possibly a white dwarf  \citep{2012ApJ...751...99T}. More similar, transient objects could be present in the known population of variable stars, but no systematic search for them has been performed thus far.

Inspired by the suggestion that SU~Lyn represent a potentially large sample of hidden symbiotic stars, \citet{2021arXiv210402686M} searched for new candidates for symbiotic stars with weak emission lines and found 33 candidates. We note that some of the stars found by \citet{2021arXiv210402686M} have accretion luminosity lower than 10~$L_\odot$. Such systems should be clearly distinguished from the classical symbiotic stars to avoid confusion in theoretical population studies, for the reasons given above. However, they represent a potentially interesting new population of systems, as was noted by \citet{2016MNRAS.461L...1M}. We note that the systems discovered by \citet{2021arXiv210402686M} are not necessarily similar to SU~Lyn, as we showed that interaction in SU~Lyn was transient, and there is no proof of transient interaction in \citet{2021arXiv210402686M} systems, with the exception of GaSS 1-31. Moreover, \citet{2021arXiv210402686M}  stated that ''\textit{the results of this paper are independent on the actual nature of the accreting star}'', which means that at least some of their systems could host a main-seaquence accretor.

\subsection{SU~Lyn as a progenitor of a classical symbiotic star}

We note that SU~Lyn may evolve towards a stable, typical symbiotic star. The mass loss of the red giant can increase significantly during the last stages of its evolution, which could result in an increased mass accretion of the red giant wind by the white dwarf on a long timescale. This could result in a mass transfer rate high enough to reach the lower limit of the accretion powered luminosity, making the system a persistent symbiotic star. However, in order to predict the future evolution of SU~Lyn orbital parameters need to be known, to predict the mass transfer mode \citep[e.g.][]{2019MNRAS.485.5468I}. We studied the future mass loss of the RG using the single star evolution models fitted to the VOSA RG parameters (Fig.~\ref{fig:SULyn_HR}). We found that in all of the models the red giant mass loss will increase by at least an order of magnitude. Assuming that the mass transfer rate would scale linearly with the mass loss from the red giant, this would imply that the mass transfer rate will increase from the current mass transfer rate  of 10$^{-10}$~M$_\odot$\,yr$^{-1}$ \citep{2018ApJ...864...46L} to mass transfer rate sufficient to make this system a classical symbiotic star.  If the increase in mass loss would be due to a He-flash, the evolution of SU~Lyn could be relatively fast \citep[e.g.][]{2007A&A...470..339M,2016ApJS..222....8D}.  The fast timescale of increase in the mass accretion rate would be similar to the evolution of CH~Cyg, which did not resemble a symbiotic star until $\sim$1960s since at least 1885 \citep[e.g.][and references therein]{1990A&A...235..219M}. Moreover, changes in the pulsation period of CH~Cyg red giant were attributed to a recent He-flash \citep{1992A&A...254..127M}. However, no direct link between the He-flash in CH~Cyg and the change in its spectroscopic appearance was made thus far, hindering a direct comparison to SU~Lyn. We predict that sufficient increase in the mass transfer rate in SU~Lyn could happen anytime between the next 10$^2$--10$^5$ years, therefore it is unlikely that it will be observed in the near future. Nonetheless, SU~Lyn could serve as an useful case study of symbiotic star progenitors.

\section{Conclusions}

SU~Lyn showed a period of high X-ray luminosity and strong UV variability, which lead \citet{2016MNRAS.461L...1M} to classify this system as a symbiotic star. However, due to lack of strong emission lines \citet{2016MNRAS.461L...1M} classified SU~Lyn as a member of a new class of hidden symbiotic stars. We showed that at the time of high X-ray luminosity the Balmer lines were as strong as in the case of classical symbiotic stars. Combined with the variable mass transfer rate it is clear that SU~Lyn can transition between being hidden and a classical symbiotic star. 

We presented optical monitoring of SU~Lyn, which revealed a strong variability in the emission lines and a weak photometric variability. The Balmer and \mbox{[O\,{\sc iii}]} lines showed strong variability that appeared to not be accompanied by any photometric variability. Among the strongest spectroscopic features was the appearance of a transient P~Cygni profile in the H$\alpha$ line. The \mbox{[O\,{\sc iii}]}~5007 line showed two variable components that followed the variability of Balmer lines. In the most recent spectra, a variable, blue-shifted absorption component of the Na~D lines is present, suggesting a strong variability in the circumstellar reddening. This is consistent with a variable source of absorption observed in the X-ray range by \citet{2018ApJ...864...46L}.

We estimated the RG effective temperature to be 3200$\pm$100~K and its luminosity to be 7700$\pm$1000~L$_\odot$. A comparison of the RG stellar parameters to the single star evolution models revealed that the RG is a low mass star (2.78$^{+0.62}_{-0.64}$ M$_\odot$) relatively early in the thermally-pulsating stage of evolution. The single star evolution models show that the RG mass loss will increase by at least an order of magnitude in the future, which suggests that SU~Lyn will evolve towards a persistent symbiotic star.

We note that  one of the main challenges with any interpretation of SU~Lyn variability is understanding whether all of the observed optical variability is associate with the X-ray flare. While it is likely that the Balmer lines were related to the X-ray flare (Fig.~\ref{fig:balmer_SOPHIE}), the variability observed in the most recent years could occur independently from the 2010--2012 X-ray event. Therefore, future monitoring and detection of possible subsequent X-ray variability will reveal the nature of the system.

\section*{Acknowledgements}

We are grateful to all the observers that made their observations available for this research. In particular, we are grateful to Paolo Berardi (BER), Christophe Boussin (CBO), Sean Curry (CUR), David Boyd (DBO), Forrest Sims (FAS), Franck Boubault (FBO), Francisco Campos (FCA), Martineau Buchet (GMAYBU), Ibrahima Diarrasouba (IBR), John P. Coffins (JCO), Joan Guarro Flo (JGF), Jacques Michelet (JMI), Jacques Montier (JMO), Jim Edlin (JPE),  	Jean-Paul Godard (JPG), Jean-Pierre Masviel (JPM), Jean-Philippe Nougayrede (JPN), James R. Foster (JRF), Kevin Gurney (KGU), Keith Shank (KSH), Tim Lester (LES), Michel Verlinden (MER), Olivier Garde (OGA), Pavol A. Dubovsk\'{y} (PAD), Paolo Cazzato (PCA), St\'{e}phane Charbonnel (SCH), Umberto Sollecchia (SOL) and Tony Rodda (TRO) for their excellent spectroscopic observations. We acknowledge with thanks the variable star observations from the AAVSO International Database contributed by observers worldwide and used in this research.

This work was supported by STFC [ST/T000244/1]. This research has been funded by the National Science Centre, Poland, through grant OPUS No. 2017/27/B/ST9/01940 and through grant K$\Pi$-06-H28/2 08.12.2018 "Binary stars with compact object" (Bulgarian National Science Fund).

Based on data retrieved from the SOPHIE archive at Observatoire de Haute-Provence (OHP), available at atlas.obs-hp.fr/SOPHIE. This publication makes use of VOSA, developed under the Spanish Virtual Observatory project supported by the Spanish MINECO through grant AyA2017-84089. VOSA has been partially updated by using funding from the European Union's Horizon 2020 Research and Innovation Programme, under Grant Agreement nº 776403 (EXOPLANETS-A). This paper makes use of data from the first public release of the WASP data (Butters et al. 2010) as provided by the WASP consortium and services at the NASA Exoplanet Archive, which is operated by the California Institute of Technology, under contract with the National Aeronautics and Space Administration under the Exoplanet Exploration Program. This work has made use of data from the European Space Agency (ESA) mission {\it Gaia} (\url{https://www.cosmos.esa.int/gaia}), processed by the {\it Gaia} Data Processing and Analysis Consortium (DPAC, \url{https://www.cosmos.esa.int/web/gaia/dpac/consortium}). Funding for the DPAC has been provided by national institutions, in particular the institutions participating in the {\it Gaia} Multilateral Agreement. The DASCH project at Harvard is grateful for partial support from NSF grants AST-0407380, AST-0909073, and AST-1313370.

\section*{DATA AVAILABILITY}
The data underlying this article is publicly available. The derived data generated in this work will be shared on a reasonable request to the corresponding author.





\bibliographystyle{mnras}
\bibliography{literature} 

\begin{thebibliography}{}
\makeatletter
\relax
\def\mn@urlcharsother{\let\do\@makeother \do\$\do\&\do\#\do\^\do\_\do\%\do\~}
\def\mn@doi{\begingroup\mn@urlcharsother \@ifnextchar [ {\mn@doi@}
  {\mn@doi@[]}}
\def\mn@doi@[#1]#2{\def\@tempa{#1}\ifx\@tempa\@empty \href
  {http://dx.doi.org/#2} {doi:#2}\else \href {http://dx.doi.org/#2} {#1}\fi
  \endgroup}
\def\mn@eprint#1#2{\mn@eprint@#1:#2::\@nil}
\def\mn@eprint@arXiv#1{\href {http://arxiv.org/abs/#1} {{\tt arXiv:#1}}}
\def\mn@eprint@dblp#1{\href {http://dblp.uni-trier.de/rec/bibtex/#1.xml}
  {dblp:#1}}
\def\mn@eprint@#1:#2:#3:#4\@nil{\def\@tempa {#1}\def\@tempb {#2}\def\@tempc
  {#3}\ifx \@tempc \@empty \let \@tempc \@tempb \let \@tempb \@tempa \fi \ifx
  \@tempb \@empty \def\@tempb {arXiv}\fi \@ifundefined
  {mn@eprint@\@tempb}{\@tempb:\@tempc}{\expandafter \expandafter \csname
  mn@eprint@\@tempb\endcsname \expandafter{\@tempc}}}

\bibitem[\protect\citeauthoryear{{Abrahamyan}, {Mickaelian}  \&
  {Knyazyan}}{{Abrahamyan} et~al.}{2015}]{2015A&C....10...99A}
{Abrahamyan} H.~V.,  {Mickaelian} A.~M.,   {Knyazyan} A.~V.,  2015, \mn@doi
  [Astronomy and Computing] {10.1016/j.ascom.2014.12.002}, \href
  {https://ui.adsabs.harvard.edu/abs/2015A&C....10...99A} {10, 99}

\bibitem[\protect\citeauthoryear{{Akras}, {Guzman-Ramirez}, {Leal-Ferreira}  \&
  {Ramos-Larios}}{{Akras} et~al.}{2019a}]{2019ApJS..240...21A}
{Akras} S.,  {Guzman-Ramirez} L.,  {Leal-Ferreira} M.~L.,   {Ramos-Larios} G.,
  2019a, \mn@doi [\apjs] {10.3847/1538-4365/aaf88c}, \href
  {https://ui.adsabs.harvard.edu/abs/2019ApJS..240...21A} {240, 21}

\bibitem[\protect\citeauthoryear{{Akras}, {Leal-Ferreira}, {Guzman-Ramirez}  \&
  {Ramos-Larios}}{{Akras} et~al.}{2019b}]{2019MNRAS.483.5077A}
{Akras} S.,  {Leal-Ferreira} M.~L.,  {Guzman-Ramirez} L.,   {Ramos-Larios} G.,
  2019b, \mn@doi [\mnras] {10.1093/mnras/sty3359}, \href
  {https://ui.adsabs.harvard.edu/abs/2019MNRAS.483.5077A} {483, 5077}

\bibitem[\protect\citeauthoryear{{Anupama}, {Kamath}, {Gurugubelli}  \&
  {Miko{\l}ajewska}}{{Anupama} et~al.}{2012}]{2012BaltA..21..172A}
{Anupama} G.~C.,  {Kamath} U.~S.,  {Gurugubelli} U.~K.,   {Miko{\l}ajewska} J.,
   2012, \mn@doi [Baltic Astronomy] {10.1515/astro-2017-0372}, \href
  {https://ui.adsabs.harvard.edu/abs/2012BaltA..21..172A} {21, 172}

\bibitem[\protect\citeauthoryear{{Bailer-Jones}, {Rybizki}, {Fouesneau},
  {Mantelet}  \& {Andrae}}{{Bailer-Jones} et~al.}{2018}]{2018AJ....156...58B}
{Bailer-Jones} C.~A.~L.,  {Rybizki} J.,  {Fouesneau} M.,  {Mantelet} G.,
  {Andrae} R.,  2018, \mn@doi [\aj] {10.3847/1538-3881/aacb21}, \href
  {https://ui.adsabs.harvard.edu/abs/2018AJ....156...58B} {156, 58}

\bibitem[\protect\citeauthoryear{{Bailer-Jones}, {Rybizki}, {Fouesneau},
  {Demleitner}  \& {Andrae}}{{Bailer-Jones} et~al.}{2021}]{2021AJ....161..147B}
{Bailer-Jones} C.~A.~L.,  {Rybizki} J.,  {Fouesneau} M.,  {Demleitner} M.,
  {Andrae} R.,  2021, \mn@doi [\aj] {10.3847/1538-3881/abd806}, \href
  {https://ui.adsabs.harvard.edu/abs/2021AJ....161..147B} {161, 147}

\bibitem[\protect\citeauthoryear{{Bayo}, {Rodrigo}, {Barrado Y Navascu{\'e}s},
  {Solano}, {Guti{\'e}rrez}, {Morales-Calder{\'o}n}  \& {Allard}}{{Bayo}
  et~al.}{2008}]{2008A&A...492..277B}
{Bayo} A.,  {Rodrigo} C.,  {Barrado Y Navascu{\'e}s} D.,  {Solano} E.,
  {Guti{\'e}rrez} R.,  {Morales-Calder{\'o}n} M.,   {Allard} F.,  2008, \mn@doi
  [\aap] {10.1051/0004-6361:200810395}, \href
  {https://ui.adsabs.harvard.edu/abs/2008A&A...492..277B} {492, 277}

\bibitem[\protect\citeauthoryear{{Belloni}, {Miko{\l}ajewska}, {I{\l}kiewicz},
  {Schreiber}, {Giersz}, {Rivera Sandoval}  \& {Rodrigues}}{{Belloni}
  et~al.}{2020}]{2020MNRAS.496.3436B}
{Belloni} D.,  {Miko{\l}ajewska} J.,  {I{\l}kiewicz} K.,  {Schreiber} M.~R.,
  {Giersz} M.,  {Rivera Sandoval} L.~E.,   {Rodrigues} C.~V.,  2020, \mn@doi
  [\mnras] {10.1093/mnras/staa1714}, \href
  {https://ui.adsabs.harvard.edu/abs/2020MNRAS.496.3436B} {496, 3436}

\bibitem[\protect\citeauthoryear{{Blind}, {Boffin}, {Berger}, {Le Bouquin},
  {M{\'e}rand}, {Lazareff}  \& {Zins}}{{Blind}
  et~al.}{2011}]{2011A&A...536A..55B}
{Blind} N.,  {Boffin} H.~M.~J.,  {Berger} J.~P.,  {Le Bouquin} J.~B.,
  {M{\'e}rand} A.,  {Lazareff} B.,   {Zins} G.,  2011, \mn@doi [\aap]
  {10.1051/0004-6361/201118036}, \href
  {https://ui.adsabs.harvard.edu/abs/2011A&A...536A..55B} {536, A55}

\bibitem[\protect\citeauthoryear{{Booth}, {Mohamed}  \&
  {Podsiadlowski}}{{Booth} et~al.}{2016}]{2016MNRAS.457..822B}
{Booth} R.~A.,  {Mohamed} S.,   {Podsiadlowski} P.,  2016, \mn@doi [\mnras]
  {10.1093/mnras/stw001}, \href
  {https://ui.adsabs.harvard.edu/abs/2016MNRAS.457..822B} {457, 822}

\bibitem[\protect\citeauthoryear{{Bouchy} \& {Sophie Team}}{{Bouchy} \& {Sophie
  Team}}{2006}]{2006tafp.conf..319B}
{Bouchy} F.,  {Sophie Team} 2006, in {Arnold} L.,  {Bouchy} F.,   {Moutou} C.,
  eds, Tenth Anniversary of 51 Peg-b: Status of and prospects for hot Jupiter
  studies. pp 319--325

\bibitem[\protect\citeauthoryear{{Bourg{\'e}s}, {Lafrasse}, {Mella},
  {Chesneau}, {Bouquin}, {Duvert}, {Chelli}  \& {Delfosse}}{{Bourg{\'e}s}
  et~al.}{2014}]{2014ASPC..485..223B}
{Bourg{\'e}s} L.,  {Lafrasse} S.,  {Mella} G.,  {Chesneau} O.,  {Bouquin}
  J.~L.,  {Duvert} G.,  {Chelli} A.,   {Delfosse} X.,  2014, in {Manset} N.,
  {Forshay} P.,  eds,  Astronomical Society of the Pacific Conference Series
  Vol. 485, Astronomical Data Analysis Software and Systems XXIII. p.~223

\bibitem[\protect\citeauthoryear{{Bujarrabal}, {Ag{\'u}ndez},
  {G{\'o}mez-Garrido}, {Kim}, {Santander-Garc{\'\i}a}, {Alcolea},
  {Castro-Carrizo}  \& {Miko{\l}ajewska}}{{Bujarrabal}
  et~al.}{2021}]{2021A&A...651A...4B}
{Bujarrabal} V.,  {Ag{\'u}ndez} M.,  {G{\'o}mez-Garrido} M.,  {Kim} H.,
  {Santander-Garc{\'\i}a} M.,  {Alcolea} J.,  {Castro-Carrizo} A.,
  {Miko{\l}ajewska} J.,  2021, \mn@doi [\aap] {10.1051/0004-6361/202141002},
  \href {https://ui.adsabs.harvard.edu/abs/2021A&A...651A...4B} {651, A4}

\bibitem[\protect\citeauthoryear{{Butters} et~al.,}{{Butters}
  et~al.}{2010}]{2010A&A...520L..10B}
{Butters} O.~W.,  et~al., 2010, \mn@doi [\aap] {10.1051/0004-6361/201015655},
  \href {https://ui.adsabs.harvard.edu/abs/2010A&A...520L..10B} {520, L10}

\bibitem[\protect\citeauthoryear{{Choi}, {Dotter}, {Conroy}, {Cantiello},
  {Paxton}  \& {Johnson}}{{Choi} et~al.}{2016}]{2016ApJ...823..102C}
{Choi} J.,  {Dotter} A.,  {Conroy} C.,  {Cantiello} M.,  {Paxton} B.,
  {Johnson} B.~D.,  2016, \mn@doi [\apj] {10.3847/0004-637X/823/2/102}, \href
  {https://ui.adsabs.harvard.edu/abs/2016ApJ...823..102C} {823, 102}

\bibitem[\protect\citeauthoryear{{Dotter}}{{Dotter}}{2016}]{2016ApJS..222....8D}
{Dotter} A.,  2016, \mn@doi [\apjs] {10.3847/0067-0049/222/1/8}, \href
  {https://ui.adsabs.harvard.edu/abs/2016ApJS..222....8D} {222, 8}

\bibitem[\protect\citeauthoryear{{Feast}, {Woolley}  \& {Yilmaz}}{{Feast}
  et~al.}{1972}]{1972MNRAS.158...23F}
{Feast} M.~W.,  {Woolley} R.,   {Yilmaz} N.,  1972, \mn@doi [\mnras]
  {10.1093/mnras/158.1.23}, \href
  {https://ui.adsabs.harvard.edu/abs/1972MNRAS.158...23F} {158, 23}

\bibitem[\protect\citeauthoryear{{Flewelling} et~al.,}{{Flewelling}
  et~al.}{2016}]{2016arXiv161205243F}
{Flewelling} H.~A.,  et~al., 2016, arXiv e-prints, \href
  {https://ui.adsabs.harvard.edu/abs/2016arXiv161205243F} {p. arXiv:1612.05243}

\bibitem[\protect\citeauthoryear{{Fox} \& {Wood}}{{Fox} \&
  {Wood}}{1982}]{1982ApJ...259..198F}
{Fox} M.~W.,  {Wood} P.~R.,  1982, \mn@doi [\apj] {10.1086/160160}, \href
  {https://ui.adsabs.harvard.edu/abs/1982ApJ...259..198F} {259, 198}

\bibitem[\protect\citeauthoryear{{Gaia Collaboration} et~al.,}{{Gaia
  Collaboration} et~al.}{2018}]{2018A&A...616A...1G}
{Gaia Collaboration} et~al., 2018, \mn@doi [\aap]
  {10.1051/0004-6361/201833051}, \href
  {https://ui.adsabs.harvard.edu/abs/2018A&A...616A...1G} {616, A1}

\bibitem[\protect\citeauthoryear{{Gaia Collaboration} et~al.,}{{Gaia
  Collaboration} et~al.}{2021}]{2021A&A...649A...1G}
{Gaia Collaboration} et~al., 2021, \mn@doi [\aap]
  {10.1051/0004-6361/202039657}, \href
  {https://ui.adsabs.harvard.edu/abs/2021A&A...649A...1G} {649, A1}

\bibitem[\protect\citeauthoryear{{Grasdalen}, {Gehrz}, {Hackwell}, {Castelaz}
  \& {Gullixson}}{{Grasdalen} et~al.}{1983}]{1983ApJS...53..413G}
{Grasdalen} G.~L.,  {Gehrz} R.~D.,  {Hackwell} J.~A.,  {Castelaz} M.,
  {Gullixson} C.,  1983, \mn@doi [\apjs] {10.1086/190897}, \href
  {https://ui.adsabs.harvard.edu/abs/1983ApJS...53..413G} {53, 413}

\bibitem[\protect\citeauthoryear{{Gromadzki}, {Miko{\l}ajewska}, {Whitelock}
  \& {Marang}}{{Gromadzki} et~al.}{2009}]{2009AcA....59..169G}
{Gromadzki} M.,  {Miko{\l}ajewska} J.,  {Whitelock} P.,   {Marang} F.,  2009,
  \actaa, \href {https://ui.adsabs.harvard.edu/abs/2009AcA....59..169G} {59,
  169}

\bibitem[\protect\citeauthoryear{{Gromadzki}, {Miko{\l}ajewska}  \&
  {Soszy{\'n}ski}}{{Gromadzki} et~al.}{2013}]{2013AcA....63..405G}
{Gromadzki} M.,  {Miko{\l}ajewska} J.,   {Soszy{\'n}ski} I.,  2013, \actaa,
  \href {https://ui.adsabs.harvard.edu/abs/2013AcA....63..405G} {63, 405}

\bibitem[\protect\citeauthoryear{{H{\o}g} et~al.,}{{H{\o}g}
  et~al.}{2000}]{2000A&A...355L..27H}
{H{\o}g} E.,  et~al., 2000, \aap, \href
  {https://ui.adsabs.harvard.edu/abs/2000A&A...355L..27H} {355, L27}

\bibitem[\protect\citeauthoryear{{I{\l}kiewicz}, {Miko{\l}ajewska}  \&
  {Monard}}{{I{\l}kiewicz} et~al.}{2017}]{2017A&A...601A.105I}
{I{\l}kiewicz} K.,  {Miko{\l}ajewska} J.,   {Monard} B.,  2017, \mn@doi [\aap]
  {10.1051/0004-6361/201630021}, \href
  {https://ui.adsabs.harvard.edu/abs/2017A&A...601A.105I} {601, A105}

\bibitem[\protect\citeauthoryear{{I{\l}kiewicz}, {Miko{\l}ajewska},
  {Belczy{\'n}ski}, {Wiktorowicz}  \& {Karczmarek}}{{I{\l}kiewicz}
  et~al.}{2019}]{2019MNRAS.485.5468I}
{I{\l}kiewicz} K.,  {Miko{\l}ajewska} J.,  {Belczy{\'n}ski} K.,  {Wiktorowicz}
  G.,   {Karczmarek} P.,  2019, \mn@doi [\mnras] {10.1093/mnras/stz760}, \href
  {https://ui.adsabs.harvard.edu/abs/2019MNRAS.485.5468I} {485, 5468}

\bibitem[\protect\citeauthoryear{{Ishihara} et~al.,}{{Ishihara}
  et~al.}{2010}]{2010A&A...514A...1I}
{Ishihara} D.,  et~al., 2010, \mn@doi [\aap] {10.1051/0004-6361/200913811},
  \href {https://ui.adsabs.harvard.edu/abs/2010A&A...514A...1I} {514, A1}

\bibitem[\protect\citeauthoryear{{Kawamuro} et~al.,}{{Kawamuro}
  et~al.}{2018}]{2018ApJS..238...32K}
{Kawamuro} T.,  et~al., 2018, \mn@doi [\apjs] {10.3847/1538-4365/aad1ef}, \href
  {https://ui.adsabs.harvard.edu/abs/2018ApJS..238...32K} {238, 32}

\bibitem[\protect\citeauthoryear{{Kenyon} \& {Webbink}}{{Kenyon} \&
  {Webbink}}{1984}]{1984ApJ...279..252K}
{Kenyon} S.~J.,  {Webbink} R.~F.,  1984, \mn@doi [\apj] {10.1086/161888}, \href
  {https://ui.adsabs.harvard.edu/abs/1984ApJ...279..252K} {279, 252}

\bibitem[\protect\citeauthoryear{{Kippenhahn}}{{Kippenhahn}}{1955}]{1955AN....282...73K}
{Kippenhahn} R.,  1955, \mn@doi [Astronomische Nachrichten]
  {10.1002/asna.19552820204}, \href
  {https://ui.adsabs.harvard.edu/abs/1955AN....282...73K} {282, 73}

\bibitem[\protect\citeauthoryear{{Kochanek} et~al.,}{{Kochanek}
  et~al.}{2017}]{2017PASP..129j4502K}
{Kochanek} C.~S.,  et~al., 2017, \mn@doi [\pasp] {10.1088/1538-3873/aa80d9},
  \href {https://ui.adsabs.harvard.edu/abs/2017PASP..129j4502K} {129, 104502}

\bibitem[\protect\citeauthoryear{{Kumar}, {Srivastava}, {Banerjee}  \&
  {Joshi}}{{Kumar} et~al.}{2021}]{2021MNRAS.500L..12K}
{Kumar} V.,  {Srivastava} M.~K.,  {Banerjee} D. P.~K.,   {Joshi} V.,  2021,
  \mn@doi [\mnras] {10.1093/mnrasl/slaa159}, \href
  {https://ui.adsabs.harvard.edu/abs/2021MNRAS.500L..12K} {500, L12}

\bibitem[\protect\citeauthoryear{{Laycock}, {Tang}, {Grindlay}, {Los}, {Simcoe}
   \& {Mink}}{{Laycock} et~al.}{2010}]{2010AJ....140.1062L}
{Laycock} S.,  {Tang} S.,  {Grindlay} J.,  {Los} E.,  {Simcoe} R.,   {Mink} D.,
   2010, \mn@doi [\aj] {10.1088/0004-6256/140/4/1062}, \href
  {https://ui.adsabs.harvard.edu/abs/2010AJ....140.1062L} {140, 1062}

\bibitem[\protect\citeauthoryear{{Lopes de Oliveira}, {Sokoloski}, {Luna},
  {Mukai}  \& {Nelson}}{{Lopes de Oliveira} et~al.}{2018}]{2018ApJ...864...46L}
{Lopes de Oliveira} R.,  {Sokoloski} J.~L.,  {Luna} G.~J.~M.,  {Mukai} K.,
  {Nelson} T.,  2018, \mn@doi [\apj] {10.3847/1538-4357/aad2d5}, \href
  {https://ui.adsabs.harvard.edu/abs/2018ApJ...864...46L} {864, 46}

\bibitem[\protect\citeauthoryear{{L{\"u}}, {Yungelson}  \& {Han}}{{L{\"u}}
  et~al.}{2006}]{2006MNRAS.372.1389L}
{L{\"u}} G.,  {Yungelson} L.,   {Han} Z.,  2006, \mn@doi [\mnras]
  {10.1111/j.1365-2966.2006.10947.x}, \href
  {https://ui.adsabs.harvard.edu/abs/2006MNRAS.372.1389L} {372, 1389}

\bibitem[\protect\citeauthoryear{{Mas-Hesse} et~al.,}{{Mas-Hesse}
  et~al.}{2003}]{2003A&A...411L.261M}
{Mas-Hesse} J.~M.,  et~al., 2003, \mn@doi [\aap] {10.1051/0004-6361:20031418},
  \href {https://ui.adsabs.harvard.edu/abs/2003A&A...411L.261M} {411, L261}

\bibitem[\protect\citeauthoryear{{Mattsson}, {H{\"o}fner}  \&
  {Herwig}}{{Mattsson} et~al.}{2007}]{2007A&A...470..339M}
{Mattsson} L.,  {H{\"o}fner} S.,   {Herwig} F.,  2007, \mn@doi [\aap]
  {10.1051/0004-6361:20066368}, \href
  {https://ui.adsabs.harvard.edu/abs/2007A&A...470..339M} {470, 339}

\bibitem[\protect\citeauthoryear{{Mikolajewska} \& {Kenyon}}{{Mikolajewska} \&
  {Kenyon}}{1992}]{1992MNRAS.256..177M}
{Mikolajewska} J.,  {Kenyon} S.~J.,  1992, \mn@doi [\mnras]
  {10.1093/mnras/256.2.177}, \href
  {https://ui.adsabs.harvard.edu/abs/1992MNRAS.256..177M} {256, 177}

\bibitem[\protect\citeauthoryear{{Mikolajewski}, {Mikolajewska}  \&
  {Khudiakova}}{{Mikolajewski} et~al.}{1990}]{1990A&A...235..219M}
{Mikolajewski} M.,  {Mikolajewska} J.,   {Khudiakova} T.~N.,  1990, \aap, \href
  {https://ui.adsabs.harvard.edu/abs/1990A&A...235..219M} {235, 219}

\bibitem[\protect\citeauthoryear{{Mikolajewski}, {Mikolajewska}  \&
  {Khudyakova}}{{Mikolajewski} et~al.}{1992}]{1992A&A...254..127M}
{Mikolajewski} M.,  {Mikolajewska} J.,   {Khudyakova} T.~N.,  1992, \aap, \href
  {https://ui.adsabs.harvard.edu/abs/1992A&A...254..127M} {254, 127}

\bibitem[\protect\citeauthoryear{{Mohamed} \& {Podsiadlowski}}{{Mohamed} \&
  {Podsiadlowski}}{2012}]{2012BaltA..21...88M}
{Mohamed} S.,  {Podsiadlowski} P.,  2012, \mn@doi [Baltic Astronomy]
  {10.1515/astro-2017-0362}, \href
  {https://ui.adsabs.harvard.edu/abs/2012BaltA..21...88M} {21, 88}

\bibitem[\protect\citeauthoryear{{Monet} et~al.,}{{Monet}
  et~al.}{2003}]{2003AJ....125..984M}
{Monet} D.~G.,  et~al., 2003, \mn@doi [\aj] {10.1086/345888}, \href
  {https://ui.adsabs.harvard.edu/abs/2003AJ....125..984M} {125, 984}

\bibitem[\protect\citeauthoryear{{Moultaka}, {Ilovaisky}, {Prugniel}  \&
  {Soubiran}}{{Moultaka} et~al.}{2004}]{2004PASP..116..693M}
{Moultaka} J.,  {Ilovaisky} S.~A.,  {Prugniel} P.,   {Soubiran} C.,  2004,
  \mn@doi [\pasp] {10.1086/422177}, \href
  {https://ui.adsabs.harvard.edu/abs/2004PASP..116..693M} {116, 693}

\bibitem[\protect\citeauthoryear{{Mukai} et~al.,}{{Mukai}
  et~al.}{2016}]{2016MNRAS.461L...1M}
{Mukai} K.,  et~al., 2016, \mn@doi [\mnras] {10.1093/mnrasl/slw087}, \href
  {https://ui.adsabs.harvard.edu/abs/2016MNRAS.461L...1M} {461, L1}

\bibitem[\protect\citeauthoryear{{Munari} \& {Zwitter}}{{Munari} \&
  {Zwitter}}{1997}]{1997A&A...318..269M}
{Munari} U.,  {Zwitter} T.,  1997, \aap, \href
  {https://ui.adsabs.harvard.edu/abs/1997A&A...318..269M} {318, 269}

\bibitem[\protect\citeauthoryear{{Munari} et~al.,}{{Munari}
  et~al.}{2021}]{2021arXiv210402686M}
{Munari} U.,  et~al., 2021, arXiv e-prints, \href
  {https://ui.adsabs.harvard.edu/abs/2021arXiv210402686M} {p. arXiv:2104.02686}

\bibitem[\protect\citeauthoryear{{Nomoto}, {Saio}, {Kato}  \&
  {Hachisu}}{{Nomoto} et~al.}{2007}]{2007ApJ...663.1269N}
{Nomoto} K.,  {Saio} H.,  {Kato} M.,   {Hachisu} I.,  2007, \mn@doi [\apj]
  {10.1086/518465}, \href
  {https://ui.adsabs.harvard.edu/abs/2007ApJ...663.1269N} {663, 1269}

\bibitem[\protect\citeauthoryear{{Pietrzy{\'n}ski} et~al.,}{{Pietrzy{\'n}ski}
  et~al.}{2019}]{2019Natur.567..200P}
{Pietrzy{\'n}ski} G.,  et~al., 2019, \mn@doi [\nat]
  {10.1038/s41586-019-0999-4}, \href
  {https://ui.adsabs.harvard.edu/abs/2019Natur.567..200P} {567, 200}

\bibitem[\protect\citeauthoryear{{Pollacco} et~al.,}{{Pollacco}
  et~al.}{2006}]{2006PASP..118.1407P}
{Pollacco} D.~L.,  et~al., 2006, \mn@doi [\pasp] {10.1086/508556}, \href
  {https://ui.adsabs.harvard.edu/abs/2006PASP..118.1407P} {118, 1407}

\bibitem[\protect\citeauthoryear{{Price}, {Smith}, {Kuchar}, {Mizuno}  \&
  {Kraemer}}{{Price} et~al.}{2010}]{2010ApJS..190..203P}
{Price} S.~D.,  {Smith} B.~J.,  {Kuchar} T.~A.,  {Mizuno} D.~R.,   {Kraemer}
  K.~E.,  2010, \mn@doi [\apjs] {10.1088/0067-0049/190/2/203}, \href
  {https://ui.adsabs.harvard.edu/abs/2010ApJS..190..203P} {190, 203}

\bibitem[\protect\citeauthoryear{{Ramsay}, {Sokoloski}, {Luna}  \&
  {Nu{\~n}ez}}{{Ramsay} et~al.}{2016}]{2016MNRAS.461.3599R}
{Ramsay} G.,  {Sokoloski} J.~L.,  {Luna} G.~J.~M.,   {Nu{\~n}ez} N.~E.,  2016,
  \mn@doi [\mnras] {10.1093/mnras/stw1546}, \href
  {https://ui.adsabs.harvard.edu/abs/2016MNRAS.461.3599R} {461, 3599}

\bibitem[\protect\citeauthoryear{{Schlafly}, {Meisner}  \& {Green}}{{Schlafly}
  et~al.}{2019}]{2019ApJS..240...30S}
{Schlafly} E.~F.,  {Meisner} A.~M.,   {Green} G.~M.,  2019, \mn@doi [\apjs]
  {10.3847/1538-4365/aafbea}, \href
  {https://ui.adsabs.harvard.edu/abs/2019ApJS..240...30S} {240, 30}

\bibitem[\protect\citeauthoryear{{Shagatova}, {Skopal}, {Shugarov},
  {Kom{\v{z}}{\'\i}k}, {Kundra}  \& {Teyssier}}{{Shagatova}
  et~al.}{2021}]{2021A&A...646A.116S}
{Shagatova} N.,  {Skopal} A.,  {Shugarov} S.~Y.,  {Kom{\v{z}}{\'\i}k} R.,
  {Kundra} E.,   {Teyssier} F.,  2021, \mn@doi [\aap]
  {10.1051/0004-6361/202039103}, \href
  {https://ui.adsabs.harvard.edu/abs/2021A&A...646A.116S} {646, A116}

\bibitem[\protect\citeauthoryear{{Shappee} et~al.,}{{Shappee}
  et~al.}{2014}]{2014ApJ...788...48S}
{Shappee} B.~J.,  et~al., 2014, \mn@doi [\apj] {10.1088/0004-637X/788/1/48},
  \href {https://ui.adsabs.harvard.edu/abs/2014ApJ...788...48S} {788, 48}

\bibitem[\protect\citeauthoryear{{Skopal} et~al.,}{{Skopal}
  et~al.}{2017}]{2017A&A...604A..48S}
{Skopal} A.,  et~al., 2017, \mn@doi [\aap] {10.1051/0004-6361/201629593}, \href
  {https://ui.adsabs.harvard.edu/abs/2017A&A...604A..48S} {604, A48}

\bibitem[\protect\citeauthoryear{{Skrutskie} et~al.,}{{Skrutskie}
  et~al.}{2006}]{2006AJ....131.1163S}
{Skrutskie} M.~F.,  et~al., 2006, \mn@doi [\aj] {10.1086/498708}, \href
  {https://ui.adsabs.harvard.edu/abs/2006AJ....131.1163S} {131, 1163}

\bibitem[\protect\citeauthoryear{{Szymczak} \& {Engels}}{{Szymczak} \&
  {Engels}}{1995}]{1995A&A...296..727S}
{Szymczak} M.,  {Engels} D.,  1995, \aap, \href
  {https://ui.adsabs.harvard.edu/abs/1995A&A...296..727S} {296, 727}

\bibitem[\protect\citeauthoryear{{Tang}, {Grindlay}, {Moe}, {Orosz}, {Kurucz},
  {Quinn}  \& {Servillat}}{{Tang} et~al.}{2012}]{2012ApJ...751...99T}
{Tang} S.,  {Grindlay} J.~E.,  {Moe} M.,  {Orosz} J.~A.,  {Kurucz} R.~L.,
  {Quinn} S.~N.,   {Servillat} M.,  2012, \mn@doi [\apj]
  {10.1088/0004-637X/751/2/99}, \href
  {https://ui.adsabs.harvard.edu/abs/2012ApJ...751...99T} {751, 99}

\bibitem[\protect\citeauthoryear{{Teyssier}}{{Teyssier}}{2019}]{2019CoSka..49..217T}
{Teyssier} F.,  2019, Contributions of the Astronomical Observatory Skalnate
  Pleso, \href {https://ui.adsabs.harvard.edu/abs/2019CoSka..49..217T} {49,
  217}

\bibitem[\protect\citeauthoryear{{Tomov}, {Stoyanov}  \& {Zamanov}}{{Tomov}
  et~al.}{2016}]{2016MNRAS.462.4435T}
{Tomov} T.~V.,  {Stoyanov} K.~A.,   {Zamanov} R.~K.,  2016, \mn@doi [\mnras]
  {10.1093/mnras/stw2012}, \href
  {https://ui.adsabs.harvard.edu/abs/2016MNRAS.462.4435T} {462, 4435}

\bibitem[\protect\citeauthoryear{{University of Virginia}}{{University of
  Virginia}}{1956}]{1956PMcCO..13a...0B}
{University of Virginia} 1956, {Publications of the Leander McCormick
  Observatory of the University of Virginia, Volume 13}

\bibitem[\protect\citeauthoryear{{Uttenthaler}, {Greimel}  \&
  {Templeton}}{{Uttenthaler} et~al.}{2016}]{2016AN....337..293U}
{Uttenthaler} S.,  {Greimel} R.,   {Templeton} M.,  2016, \mn@doi
  [Astronomische Nachrichten] {10.1002/asna.201512296}, \href
  {https://ui.adsabs.harvard.edu/abs/2016AN....337..293U} {337, 293}

\bibitem[\protect\citeauthoryear{{Verbunt} \& {Rappaport}}{{Verbunt} \&
  {Rappaport}}{1988}]{1988ApJ...332..193V}
{Verbunt} F.,  {Rappaport} S.,  1988, \mn@doi [\apj] {10.1086/166645}, \href
  {https://ui.adsabs.harvard.edu/abs/1988ApJ...332..193V} {332, 193}

\bibitem[\protect\citeauthoryear{{Voges} et~al.,}{{Voges}
  et~al.}{1999}]{1999A&A...349..389V}
{Voges} W.,  et~al., 1999, \aap, \href
  {https://ui.adsabs.harvard.edu/abs/1999A&A...349..389V} {349, 389}

\bibitem[\protect\citeauthoryear{{Wood}}{{Wood}}{2015}]{2015MNRAS.448.3829W}
{Wood} P.~R.,  2015, \mn@doi [\mnras] {10.1093/mnras/stv289}, \href
  {https://ui.adsabs.harvard.edu/abs/2015MNRAS.448.3829W} {448, 3829}

\makeatother
\end{thebibliography}



\appendix

\clearpage
\onecolumn

\section{Log of spectroscopic observations}

\begin{center}
\begin{longtable}{ccccccccc}
\caption{ List of spectroscopic observations, equivalent widths of emission lines and equivalent widths of the circumbinary absorption component of the Na~D lines. The list of ARAS observers and the site names can be found at the ARAS website (https://aras-database.github.io/database/symbiotics.html). }\label{logspec}\\
\hline	
JD-2450000	&	Observer	&	Site	&	EW(H$\alpha$) [\AA]	&	EW(H$\beta$) [\AA]	&	EW(\mbox{[O\,{\sc iii}]}~5007) [\AA]	&	EW(Na D1) [\AA]	&	EW(Na D2) [\AA]	\\\hline															
5999.30	&		&	Sophie	&	6.4	&	4.0	&		&	0.03	&	0.09	\\
7513.39	&	FTE	&	ROU-FR	&	2.3	&	2.7	&	1.4	&	0.22	&	0.30	\\
7775.28	&	FTE	&	ROU-FR	&	1.0	&	1.1	&	0.4	&	0.29	&	0.44	\\
7801.30	&	FTE	&	ROU-FR	&	0.7	&	1.1	&	0.2	&	0.32	&	0.48	\\
7840.35	&	FTE	&	ROU-FR	&	1.3	&	1.1	&	0.4	&	0.35	&	0.51	\\
7841.43	&	OGA	&	OTO-FR	&	1.2	&	0.9	&	0.5	&	0.40	&	0.52	\\
7842.33	&	JGF	&	PIE-SP	&	1.1	&	1.1	&	0.7	&	0.39	&	0.53	\\
7846.34	&	FTE	&	ROU-FR	&	1.4	&	1.3	&	0.6	&	0.36	&	0.54	\\
7847.46	&	JGF	&	PIE-SP	&	1.9	&	1.6	&	0.8	&	0.37	&	0.53	\\
7850.44	&	JGF	&	PIE-SP	&	1.8	&	1.6	&	0.8	&	0.35	&	0.48	\\
7856.43	&	JGF	&	PIE-SP	&	1.7	&	1.6	&	0.7	&	0.33	&	0.49	\\
7864.35	&	OGA	&	OTO-FR	&	1.9	&	1.5	&	1.0	&	0.36	&	0.49	\\
7864.38	&	FTE	&	ROU-FR	&	2.0	&	1.6	&	0.9	&	0.36	&	0.49	\\
7883.36	&	OGA	&	OTO-FR	&	1.7	&	1.3	&	0.9	&	0.35	&	0.51	\\
7884.37	&	FTE	&	ROU-FR	&	1.3	&	1.3	&	0.6	&	0.37	&	0.51	\\
8029.67	&	FTE	&	ROU-FR	&	1.2	&	1.2	&	1.0	&	0.35	&	0.50	\\
8058.70	&	FTE	&	ROU-FR	&	1.3	&	1.5	&	1.1	&	0.33	&	0.51	\\
8061.71	&	LES	&	MRO-CA	&	1.3	&	1.3	&	1.0	&	0.30	&	0.46	\\
8130.45	&	JGF	&	PIE-SP	&	1.3	&	1.7	&	1.4	&	0.33	&	0.46	\\
8140.36	&	JGF	&	SMM-SP	&	1.2	&	1.6	&	1.4	&	0.28	&	0.48	\\
8141.47	&	JGF	&	SMM-SP	&	1.1	&	1.7	&	1.5	&	0.31	&	0.44	\\
8142.38	&	OGA	&	OTO-FR	&	1.1	&	1.6	&	1.4	&	0.30	&	0.49	\\
8142.47	&	JGF	&	PIE-SP	&	1.2	&	1.7	&	1.2	&	0.34	&	0.49	\\
8152.48	&	JGF	&	PIE-SP	&	1.3	&	1.8	&	1.6	&	0.33	&	0.48	\\
8154.32	&	FTE	&	ROU-FR	&	1.6	&	1.9	&	1.9	&	0.32	&	0.49	\\
8157.43	&	JGF	&	PIE-SP	&	1.7	&	2.0	&	1.9	&	0.32	&	0.44	\\
8162.34	&	OGA	&	OTO-FR	&	1.3	&	1.9	&	1.7	&	0.29	&	0.43	\\
8164.58	&	LES	&	MRO-CA	&	1.2	&	1.5	&	1.4	&	0.33	&	0.42	\\
8173.29	&	FTE	&	ROU-FR	&	1.8	&	2.0	&	1.6	&	0.33	&	0.43	\\
8174.47	&	JGF	&	SMM-SP	&	1.3	&	1.6	&	1.1	&	0.34	&	0.44	\\
8186.34	&	FTE	&	ROU-FR	&	1.5	&	1.7	&	1.7	&	0.30	&	0.43	\\
8198.33	&	FTE	&	ROU-FR	&	0.9	&	1.2	&	1.2	&	0.31	&	0.44	\\
8198.43	&	JGF	&	SMM-SP	&	1.0	&	0.8	&	1.0	&	0.32	&	0.45	\\
8229.59	&	LES	&	MRO-CA	&	1.2	&	1.2	&	1.9	&	0.31	&	0.44	\\
8251.60	&	LES	&	MRO-CA	&	1.1	&	1.4	&	1.8	&	0.30	&	0.43	\\
8510.41	&	JGF	&	SMM-SP	&	1.1	&	1.6	&	1.4	&	0.28	&	0.40	\\
8518.31	&	FTE	&	ROU-FR	&	0.8	&	1.6	&	1.0	&	0.26	&	0.41	\\
8524.49	&	JGF	&	SMM-SP	&	1.4	&	1.8	&	1.9	&	0.25	&	0.41	\\
8528.30	&	FTE	&	ROU-FR	&	1.5	&	1.8	&	2.2	&	0.27	&	0.40	\\
8531.44	&	JGF	&	SMM-SP	&	1.2	&	1.9	&	1.9	&	0.26	&	0.38	\\
8540.35	&	FTE	&	ROU-FR	&	1.3	&	2.0	&	2.3	&	0.27	&	0.42	\\
8545.42	&	JGF	&	SMM-SP	&	1.3	&	2.1	&	2.1	&	0.26	&	0.38	\\
8550.31	&	FTE	&	ROU-FR	&	1.5	&	2.1	&	2.1	&	0.25	&	0.39	\\
8551.67	&	LES	&	MRO-CA	&	1.3	&	2.0	&	1.9	&	0.27	&	0.38	\\
8553.34	&	JGF	&	SMM-SP	&	1.4	&	2.1	&	1.9	&	0.24	&	0.38	\\
8563.32	&	OGA	&	OTO-FR	&	1.1	&	1.7	&	1.4	&	0.25	&	0.39	\\
8566.56	&	LES	&	MRO-CA	&	0.9	&	1.4	&	1.4	&	0.27	&	0.39	\\
8568.36	&	JGF	&	PIE-SP	&	0.9	&	1.5	&	1.4	&	0.25	&	0.38	\\
8574.36	&	SCH	&	DUR-FR	&	1.0	&	1.5	&	1.4	&	0.24	&	0.39	\\
8585.37	&	SCH	&	DUR-FR	&	0.8	&	1.2	&	0.9	&	0.26	&	0.38	\\
8587.45	&	JGF	&	SMM-SP	&	1.0	&	1.4	&	1.2	&	0.25	&	0.38	\\
8594.38	&	SCH	&	DUR-FR	&	1.1	&	1.4	&	1.3	&	0.20	&	0.39	\\
8595.58	&	LES	&	MRO-CA	&	1.1	&	1.4	&	1.3	&	0.22	&	0.40	\\
8603.38	&	FTE	&	ROU-FR	&	1.1	&	1.3	&	1.3	&	0.22	&	0.38	\\
8819.39	&	FTE	&	SMM-SP	&	0.8	&	1.3	&	1.5	&	0.22	&	0.34	\\
8848.41	&	OGA	&	OTO-FR	&	0.9	&	1.5	&	1.2	&	0.19	&	0.36	\\
8861.31	&	FTE	&	ROU-FR	&	1.0	&	1.5	&	1.3	&	0.20	&	0.37	\\
8885.46	&	SCH	&	DUR-FR	&	0.7	&	1.2	&	1.0	&	0.21	&	0.35	\\
8895.32	&	FTE	&	SMM-SP	&	0.8	&	1.4	&	1.1	&	0.18	&	0.30	\\
8923.34	&	FTE	&	ROU-FR	&	1.1	&	1.2	&	1.0	&	0.18	&	0.36	\\
8933.59	&	LES	&	MRO-CA	&	1.3	&	1.5	&	1.0	&	0.23	&	0.40	\\
8935.31	&	FTE	&	ROU-FR	&	1.1	&	1.4	&	0.9	&	0.25	&	0.39	\\
8955.36	&	FTE	&	ROU-FR	&	0.9	&	1.4	&	0.7	&	0.20	&	0.39	\\
8962.60	&	LES	&	MRO-CA	&	1.0	&	1.4	&	1.0	&	0.23	&	0.40	\\
8964.62	&	LES	&	MRO-CA	&	1.1	&	1.5	&	0.9	&	0.19	&	0.38	\\
9111.82	&	LES	&	MRO-CA	&	0.8	&	0.9	&	0.1	&	0.29	&	0.45	\\
9162.72	&	LES	&	MRO-CA	&	0.7	&	1.1	&	0.5	&	0.29	&	0.47	\\
9242.43	&	JGF	&	PIE-SP	&	0.7	&	0.8	&	0.7	&	0.20	&	0.37	\\
9273.32	&	FTE	&	ROU-FR	&	0.7	&	0.6	&	0.5	&	0.18	&	0.36	\\
9289.35	&	JGF	&	SMM-SP	&	0.8	&	0.7	&	0.7	&	0.16	&	0.31	\\
9294.45	&	JGF	&	SMM-SP	&	0.8	&	1.1	&	0.7	&	0.19	&	0.36	\\\hline																											
\end{longtable}
\end{center}

\clearpage
\twocolumn



\bsp	
\label{lastpage}
\end{document}